\documentclass[final,3p,times,twocolumn]{elsarticle}

\usepackage{amssymb}
\usepackage[colorlinks=true, allcolors=blue]{hyperref}

\usepackage{amsmath}

\journal{Applied Surface Science}

\DeclareUnicodeCharacter{2212}{\textminus}
\DeclareUnicodeCharacter{2215}{\textminus}
\begin{document}

\begin{frontmatter}

\title{Selective faceting induced by Hexagonal Boron Nitride on curved Platinum substrate}

\author[1,2,3]{Alaa Mohammed Idris Bakhit\corref{cor1}}
\author[4]{Muhammad Saad Mahmood}
\author[5,6]{Khadiza Ali}
\author[7]{Anna A. Makarova}
\author[8,9]{Igor Piš}
\author[9]{Federica Bondino}
\author[1]{Frederik Schiller}
\author[4]{Maria Peressi\corref{cor2}}

\address[1]{Centro de F\'{\i}sica de Materiales (CSIC-UPV-EHU) and Materials Physics Center (MPC), E-20018 San Sebasti\'an, Spain}
\address[2]{Departamento de F\'{\i}sica Aplicada I, Universidad del Pa\'{\i}s Vasco UPV/EHU, 20018 San Sebasti\'an, Spain}
\address[3]{Donostia International Physics Center, 20018 Donostia-San Sebasti\'an, Spain.}
\address[4]{Department of Physics,  University of Trieste,  Via Alfonso Valerio, 2, 34127 Trieste, Italy.}
\address[5]{Department of Physics, BITS Pilani, Hyderabad Campus, Hyderabad, 500078, Telangana, India}
\address[6]{Chalmers University of Technology, Göteborg, Chalmersplatsen 4, 412 96 Göteborg, Sweden}
\address[7]{Physikalische Chemie, Institut für Chemie und Biochemie, Freie Universität Berlin, Arnimallee 22, 14195 Berlin, Germany}
\address[8]{Institute of Electrical Engineering, Slovak Academy of Sciences, Dúbravská cesta 9, 84104 Bratislava, Slovakia}
\address[9]{CNR - Istituto Officina dei Materiali (IOM), S.S. 14 km 163.5, 34149 Basovizza, Trieste, Italy}

\cortext[cor1]{alaa.mohammed$@$dipc.org}
\cortext[cor2]{peressi$@$units.it}

\begin{abstract}

This work combines experimental investigations and first-principles calculations to study hexagonal boron nitride (hBN) growth on a curved Pt(331) substrate. Low-energy electron diffraction (LEED) and scanning tunnelling microscopy (STM) were employed to investigate the structural properties, observing that hBN growth led to faceting with several stable facets: Pt(111), Pt(221), Pt(441), Pt(881), and Pt(110). Additionally, the electronic properties were examined by X-ray photoemission spectroscopy (XPS) and near-edge X-ray absorption fine structure (NEXAFS), revealing an increase in the hBN-substrate interaction strength, from the weakest at the Pt(111) to the strongest at the Pt(110). Density functional theory (DFT) calculations confirm that the selective faceting mechanism is induced by hBN, which changes the order of relative stability of the uncovered facets. Although a complete understanding of the process is not achievable using only thermodynamical arguments, the comparison between hBN adhesion energies and clean surface free energies clearly indicates that some facets are stabilized with respect to others by stronger overlayer-substrate interaction, whereas for others the clean surface energetics plays the dominant role.
\end{abstract}


\end{frontmatter}

\section{Introduction}\label{sec1}
The breakthrough of graphene discovery \cite{novoselov2004electric} has sparked a revolution in materials science, leading to the exploration of a diverse family of two-dimensional (2D) materials. Among these, hexagonal boron nitride (hBN) \cite{Watanabe2004, Cassabois2016} is an isostructural, electrically insulating counterpart to graphene \cite{Geim2007}. This material is of technological interest as a substrate for other 2D materials, a protective layer \cite{bakhit2023ferromagnetic,bakhit2026formation}, or an insulating layer for electronics and spintronics applications, among others \cite{dean2010, ferrari2015science}. 
\begin{figure}[t]
    \centering
    \includegraphics[scale=0.5]{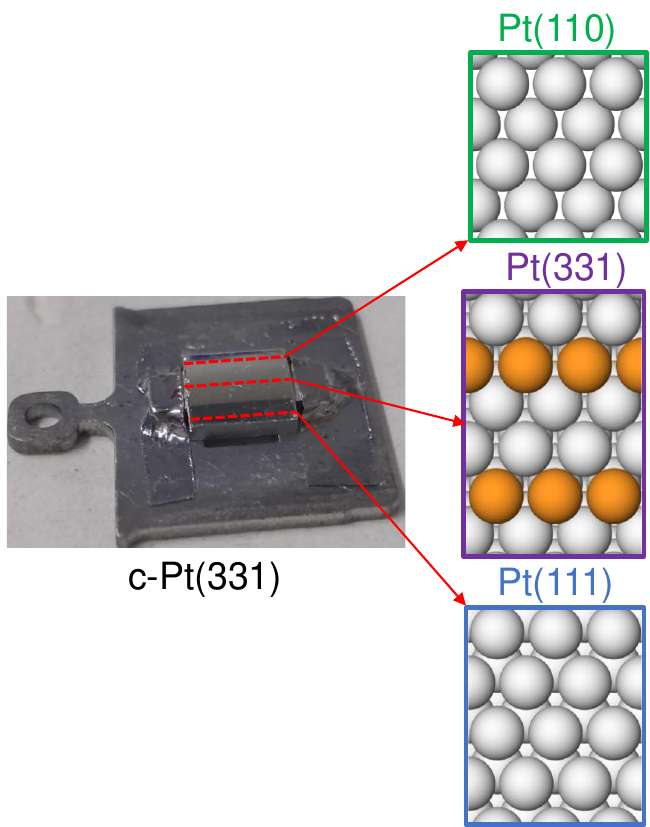}
    \caption{Curved Pt(331) crystal employed for this investigations with Pt(111), Pt(331) and Pt(110) surfaces indicated.}
    \label{fig:1}
\end{figure}
Achieving high-quality crystalline graphene or hBN for applications requires a clean, well-defined interface with other materials. Therefore, for the case of hBN, the monolayer (ML) grown on transition metal (TM) substrates has been extensively studied~\cite{auwarter2019hexagonal, pakdel2014nano, laskowski2008bonding}, revealing two key factors that control the growth: (i) the substrate type as reported by Bokdam et al.~\cite{Bokdam2014} and (ii) the lattice mismatch between hBN and the metal support. For instance, in hBN/Ni(111), hBN has an almost perfect lattice match and strong interfacial interaction with Ni(111), facilitating epitaxial growth in which hBN aligns directly with the substrate lattice, yielding a commensurate (1$\times$1) structure~\cite{nagashima1995electronic, fernandez2019boron}. For lattice-mismatch systems, on the one hand, hBN forms a self-assembled nanostructure known as "nanomesh" on Rh(111), which was first demonstrated by Corso et al.~\cite{corso2004boron} and subsequently observed also on Ru(0001) \cite{goriachko2007self}. On the other hand, Moiré structures have been reported for hBN grown on Pd(111)~\cite{morscher2006formation, nagashima1995electronic}, Ir(111)~\cite{schulz2014epitaxial}, and Pt(111)~\cite{preobrajenski2007influence}. In addition to experiments, theoretical investigations have been conducted extensively to uncover the structural properties of hBN monolayers on various flat transition metal substrates~\cite{Grad_2003, laskowski2010ab, Arias_2018, farwick_2016}. DFT simulations reveal that the strength and nature of the hBN–substrate interactions vary significantly, influencing the resulting surface morphology.

Beyond structural properties, the bonding nature at the hBN/metal interface ranges from weak van der Waals interactions to stronger chemical bonding. This leads to a significant alteration in the electronic properties of the hBN monolayer~\cite{gomez2013hexagonal}. Despite that, hBN growth has been predominantly studied on low-index single-crystal surfaces such as (111) and (110)~\cite{achilli2018growth, preobrajenski2007influence}. However, real-world substrates, such as copper foils commonly used for graphene and hBN synthesis, inevitably contain steps, defects, and additional facets beyond ideal low-index faces. The influence of stepped surfaces has been investigated for graphene growth on vicinal surfaces~\cite{casarin2014thinnest, vsrut2013temperature}. It is worth emphasizing that hBN-metal interactions at such interfacial sites extend beyond structural considerations and can be intentionally used to functionalize hBN. One prominent example is the oxygen reduction reaction (ORR), in which the metal substrate has been shown to substantially enhance the intrinsic catalytic activity of the hBN monolayer~\cite{lyalin2013functionalization}, emphasizing the broader potential of interface engineering in 2D materials. Consequently, extending the understanding and precise control of hBN-metal interactions beyond conventional low-index surfaces towards the structural complexity of stepped substrates is crucial for tuning the properties of 2D materials and enabling their use in future technological applications.

Moreover, stepped surfaces play a far more active role in epitaxial growth than their use as simple structural templates might suggest. Upon overlayer deposition, vicinal surfaces frequently undergo substantial structural reconstruction, manifesting as mound formation~\cite{Johnson1994}, step bunching, or nano- and mesoscale faceting~\cite{HORNVONHOEGEN1999475,BACHMANN2003L143}. Among these, adsorbate-induced faceting represents the most dramatic response, in which the initially uniform step distribution transforms into a periodic arrangement of alternating facets with distinct crystallographic orientations~\cite{BACHMANN2003L143}. This transformation is governed by the interplay between low surface free energy~\cite{Nien1999, Herring1951}, surface stress~\cite{WEBB1994454}, step energy~\cite{Qian1986, Chadi1987}, step–step interactions~\cite{Pehlke1991}, and the chemical coupling between the overlayer and the substrate. Theoretically, elastic continuum theory predicts that stress energy differences between coexisting facet phases give rise to long-range interactions that stabilize periodic domain patterns at the mesoscopic scale~\cite{BACHMANN2003L143}. Nevertheless, adsorbate-induced faceting remains considerably more complex than the spontaneous faceting of clean surfaces, owing to large-scale mass transport and the simultaneous restructuring of two chemically distinct species, making the process inherently sensitive to kinetic limitations. Signatures of such behavior have already been reported in recent investigations of hBN on stepped Ni(111)~\cite{fernandez2019boron}, Rh(337)~\cite{ali2021atomically}, and Pt(111)~\cite{bakhit2022structure}, and have demonstrated that hBN growth on vicinal surfaces can lead to significant modification of the underlying vicinal surface, including pronounced substrate faceting. These findings highlight the need for a more comprehensive systematic investigation of hBN–metal interactions on stepped surfaces, which motivates the present work. Therefore, curved single crystals~\cite{Corso2009, Ortega2011, Max2017, Walter2015, Ortega2013, Ortega2016} ( e.g. c-Pt(331) shown in Fig. \ref{fig:1}) offer the most direct route to addressing this challenge, as their continuously varying miscut angle provides simultaneous access to a broad range of step densities and vicinal orientations within a single sample and under identical experimental conditions, enabling a truly systematic investigation of faceting phenomena in the hBN–metal system.

A comprehensive analysis of how the surface structural and electronic properties evolve with variations in step density and terrace width is achieved using advanced microscopy and spectroscopy techniques. Specifically, here, structural and electronic properties of hBN growth on a curved Pt(331) substrate were experimentally characterized using techniques such as low-energy electron diffraction (LEED), scanning tunneling microscopy (STM), X-ray photoelectron spectroscopy (XPS), near-edge X-ray absorption fine structure (NEXAFS), and angle-resolved photoemission spectroscopy (ARPES). The results show that hBN growth leads to the formation of stable facets, namely Pt(111), Pt(221), Pt(441), Pt(881), and Pt(110). To further understand the relative stability of these facets, first-principles density functional theory (DFT) calculations were performed, carefully considering the strain state induced in hBN by the overlayer-substrate mismatch. Both experimentally observed stable and unstable facets were analyzed, revealing that, in general, the former show slightly stronger adhesion to the overlayer than the latter. However, this cannot be considered the only origin of the faceting mechanism, which, as DFT results confirm, results from a complex interplay between many factors. Based on DFT calculations, we also provide a rationale for a preferential alignment of the hBN zig-zag directions with the $\langle 110\rangle$ directions and hence with the step edges. Here, these theoretical insights are consistent with our experimental findings.

\begin{figure*}[ht!]
    \centering
    \includegraphics[scale=0.53]{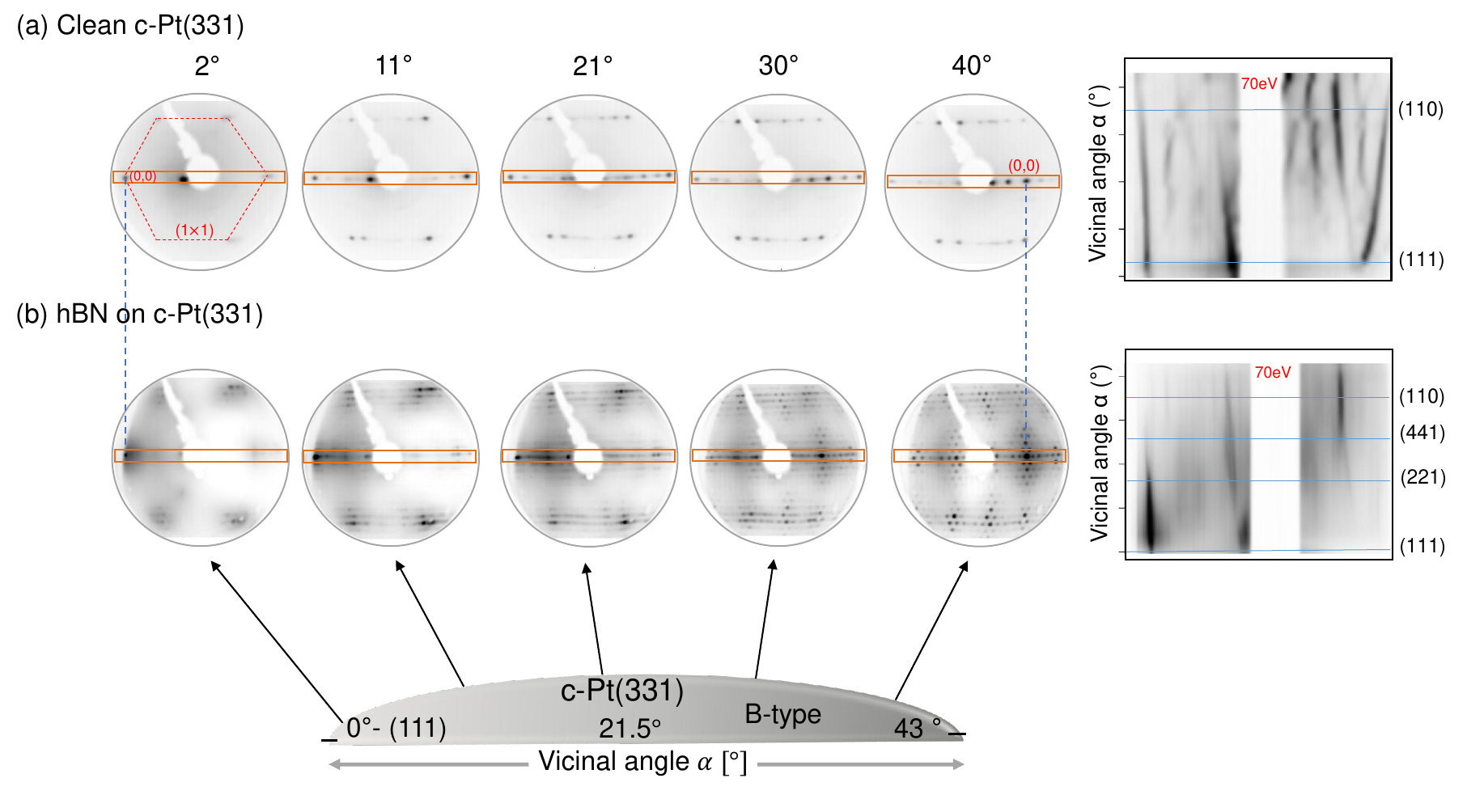}
    \caption{LEED images (a) clean c-Pt(331) and (b) hBN grown on the c-Pt(331) as a function of the vicinal angle (2$^{\circ}$, 11$^{\circ}$, 21$^{\circ}$, 30$^{\circ}$ and 40$^{\circ}$). The black arrows indicate the positions where the LEED images were taken along the curved crystal. The right-hand images display the intensity variations of the central horizontal orange rectangle in the LEED images as a function of the vicinal angle. LEED images were captured at a kinetic energy of 70 eV~\cite{bakhit2022structure}.}
    \label{fig:2}
\end{figure*}

\section{Methods}\label{sec2}
\subsection{Experimental methods}\label{sec2.1}
 
The curved c-Pt(331) single crystal was cleaned via several cycles of Argon ion sputtering (1.0 kV) at a pressure of 3 $\cdot$ $10^{-6}$ mbar and at room temperature, then annealed to 950 K. The sample was heated in an oxygen atmosphere, occasionally at 5 $\cdot$ $10^{-8}$ mbar to remove carbon impurities, maintaining a temperature of 1100 K for 5 minutes, and lastly flash-annealed to 1200 K for 4 minutes. Fig. \ref{fig:2}, a c-Pt(331) has a length of 11 mm and a diameter of 15 mm and spans a vicinal angle range of 0.0,...,43.0$^{\circ}$. This curved crystal contains only the B-step type with the (111)-0.0$^{\circ}$ surface at its edge and the (110)-35$^{\circ}$ surface close to the other edge. 

The synthesis of hBN on curved Pt(331) was achieved using the Chemical Vapor Deposition (CVD) technique. For this purpose, a clean c-Pt(331) substrate was heated and exposed to the borazine precursor (B$_{3}$H$_{6}$N$_{3}$, KATCHEM spol. s r.o.) for 15 minutes at a pressure of 5 $\cdot$ $10^{-7}$ mbar. The substrate was maintained at 1020 K, resulting in the growth of a uniform hBN monolayer across the entire curvature. 
LEED and STM experiments were conducted under ultra-high vacuum (UHV) conditions at the Material Physics Center (MPC) in San Sebastian, Spain, maintaining a base pressure of 5 $\cdot$ $10^{-10}$ mbar. LEED measurements utilized the conventional three-grid design from Omicron equipment. STM measurements were conducted in constant current mode using an Omicron VT-STM microscope, with image analysis processed via the WSXM program \cite{horcas2007wsxm}. Both techniques were performed at room temperature. 
ARPES, NEXAFS and XPS investigations were conducted at the BACH beamline of the Elettra synchrotron (Trieste, Italy). At the Elettra synchrotron, p-polarized light was used with a VG-Scienta hemispherical electron analyzer (CNR) at an angle of 60$^{\circ}$ relative to the X-ray incidence direction. The beam size on the sample was 230$\times$100 $\mu$m$^{2}$. The ARPES spectra were collected with an angular resolution of $\Delta \theta$ = 1$^{\circ}$ and an energy resolution of 30 meV. The XPS spectra were performed using a photon energy of 480 eV. The NEXAFS spectra were acquired in partial electron-yield mode, recording B KLL and N KLL Auger electrons with the electron analyzer fixed at kinetic energies of 167 and 380 eV, respectively. The intensities were normalized to the photon beam flux, measured by the drain current on a reference golden grid. The photon energy scale was calibrated with an accuracy of 0.2 eV. Furthermore, initial photoemission spectroscopy experiments were conducted at the GELEM dipole beamline of the BESSY II electron storage ring, operated by the Helmholtz-Zentrum Berlin für Materialien und Energie in Berlin, Germany.

\subsection{Computational approach}\label{sec2.2}

The calculations were carried out using the Quantum ESPRESSO ab initio simulation package~\cite{Giannozzi_2009, Giannozzi_2017, Giannozzi_2020}, based on a plane-wave basis set framework that expands the wave functions and charge density. Garrity-Bennett-Rabe-Vanderbilt (GBRV) ultrasoft pseudopotentials~\cite{garrity2014pseudopotentials} 
with the Perdew-Burke-Ernzerhof (PBE) form of exchange-correlation~\cite{Perdew_1996,ernzerhof1999assessment} have been used, with  
DFT-D3 correction added for the dispersion forces~\cite{grimme2010consistent}
Kinetic energy cutoffs of 80 Ry for the wavefunctions and 640 Ry for the charge density have been used, ensuring computational reliability and accuracy (see Supplementary Material (SM)~\cite{SM}). Supercells with slab geometry (from 4 to 7 layers of Pt, corresponding to a thickness from 6 to 9 {\AA}) have been considered for the different facet orientations, with appropriate in-plane dimensions that are dictated by the Pt substrate, while hBN is accommodated with a strain generally smaller than 3$\%$. The vacuum space between the repeated images is about 15-20 {\AA}. For hBN and Pt bulk calculations, k-point meshes of $6\times 6\times 1$ and  $6\times 6\times 6$ have been used, respectively. The supercell size is such that the ${\Gamma}$ point alone is enough for an efficient sampling of the Brillouin zone (BZ). Other structural details of the supercells are reported in SM~\cite{SM}. 
The structures are optimised by fixing the two or three lowermost Pt layers, while all remaining Pt atoms and the hBN layer are allowed to relax until the maximum atomic forces are less than 0.01 eV$\cdot$\AA$^{-1}$ and the total energy is converged within 10$^{-3}$ eV per atom, approximately.
All calculations were executed using Quantum Espresso installed on the HPC cluster LEONARDO (CINECA, Bologna).

\section{Results and discussion}\label{sec:3} 

\subsection{The growth and structural properties of hBN/c-Pt(331) system}\label{sec3.1}
A comprehensive structural characterization was carried out using LEED both prior to and after hBN growth on a curved c-Pt(331) substrate. In Fig. \ref{fig:2}(a), a hexagonal LEED pattern is observed close to the (111) at a vicinal angle of 2$^{\circ}$ due to the presence of large terraces of (111), demonstrating the three-fold symmetry characteristic of a pure face-centred cubic (fcc) of Pt(111) surface, indicating the presence of large terraces of (111). As a result of the emergence of steps when moving along the curvature, the (111) terrace width shrinks. This led to spot splitting becoming evident, resulting from the diffraction of the step array in the direction perpendicular to the steps. This splitting seen at vicinal angles 11$^{\circ}$, 21$^{\circ}$, 30$^{\circ}$ and 40$^{\circ}$ corresponds to $2\pi/d$ in reciprocal space, where $d$ is the distance from one step edge to the very next step edge. At a critical vicinal angle, the (111) terraces completely vanish, making way for (110) terraces that manifest as a rectangular LEED pattern in Fig. \ref{fig:2} (a) from 30$^{\circ}$ vicinal angle and beyond. 

Following the CVD growth of hBN, significant structural modifications were observed, as seen in Fig. \ref{fig:2} (b), corresponding to the hBN/c-Pt(331) system. On Pt(111), as has been reported by Preobrajenski et al.~\cite{preobrajenski2007influence}, due to the lattice mismatch, a coincidence lattice is found: (10 $\times$ 10) hBN unit cells on a (9 $\times$ 9) Pt atoms. In panel (b), a  Moir{\'e} pattern was detected near the (111) surface at a vicinal angle of 2$^{\circ}$. The observed structure might be slightly different from the Moir{\'e} pattern at (111). As the vicinal angle increases from 11$^{\circ}$ to 40$^{\circ}$, the surface undergoes a distinct transformation, suggesting a reconstruction process and faceting. This phenomenon was induced by hBN growth when grown on curved Ni(111), as reported by Fernandez et al.~\cite{fernandez2019boron}. Moreover, the images on the right-hand side highlight intensity variations, particularly within the central horizontal orange rectangle in the LEED images. The top image reveals crossing features attributed to the continuous appearance of steps along the clean curved crystal as a result of the terrace width shrinking; however, parallel features in panel (b) indicate the emergence of stable facet formation. The schematic diagram provides a detailed representation of the c-Pt(331) crystal, offering a clearer visualisation of the surface evolution. 

\begin{figure}[t]
    \centering
    \includegraphics[scale=0.33]{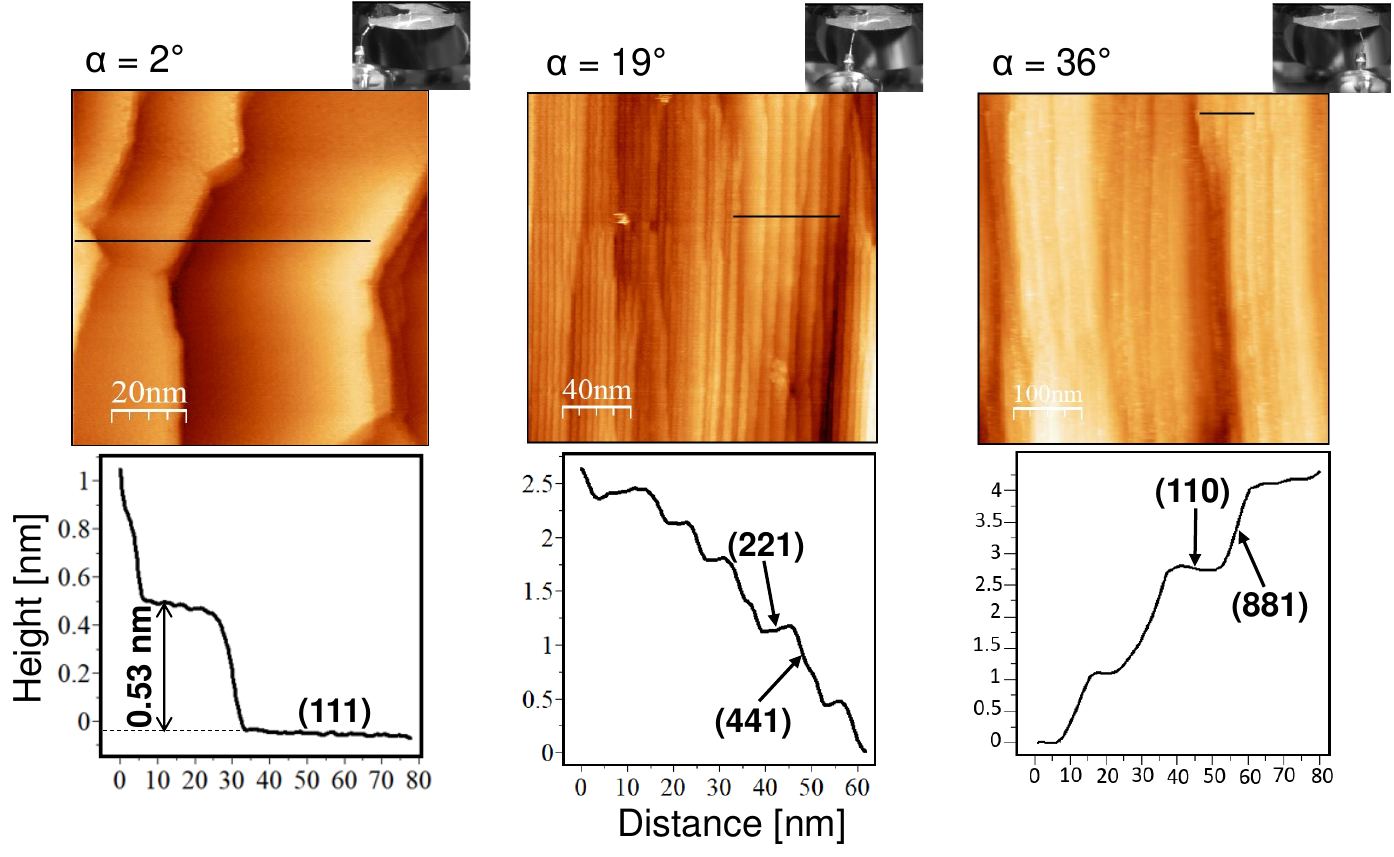}
    \caption{STM images of hBN-covered c-Pt(331) at different positions of the curved substrate ($I$=2 nA; $U_{\text{bais}}$=1 V). The bottom panels indicate the height profiles for the indicated black lines in the STM images~\cite{bakhit2022structure}.}
    \label{fig:3}
\end{figure}

Furthermore, the surface morphology after hBN growth on the c-Pt(331) system was investigated using STM topography shown in Fig. \ref{fig:3}. STM images were captured at three different positions along the curved crystal and exhibited significant morphological variations, with a distinct step height of approx. 0.50 nm observed at 2$^{\circ}$ vicinal angle, suggesting surface faceting. Notably, a clean c-Pt(331) surface exhibits a monatomic step height of 2.26 ~\AA \cite{ma8062935}. Moving towards the (331) surface at a vicinal angle of 21$^{\circ}$, a newly formed facet structure appears. Here, (221)-$16^{\circ}$ and (441)-$25^{\circ}$ were found to be the next stable facets. This means that all vicinal surfaces between (221)-$16^{\circ}$ and (441)-$25^{\circ}$ will decay into these two stable facets. Moreover, near the (110)-$35^{\circ}$ surface, two distinct facets, (881)-$30.21^{\circ}$ and (110)-$35^{\circ}$, emerge, providing additional evidence of surface restructuring. Previously, in the results reported by us~\cite{bakhit2022structure}, the (991)-$30.77^{\circ}$ facet was misidentified. This misassignment arose from the close similarity between the two facets, as reflected in their respective miscut angle values, which fall within our estimated uncertainty of approximately $\pm 0.5^\circ$.

To summarise, in the hBN/c-Pt(331) system between (111) and (110), we determined five stable facets: Pt(111), Pt(221), Pt(441), Pt(881), Pt(110). These findings were consistent with the LEED-IV analysis presented in Fig. S1 in the SM~\cite{SM}, where the stable facets Pt(111), Pt(221), and Pt(110) were identified.

\subsection{Electronic properties of hBN/c-Pt(331)
system}\label{sec3.2}
\begin{figure}[ht!]
    \centering
    \includegraphics[scale=0.4]{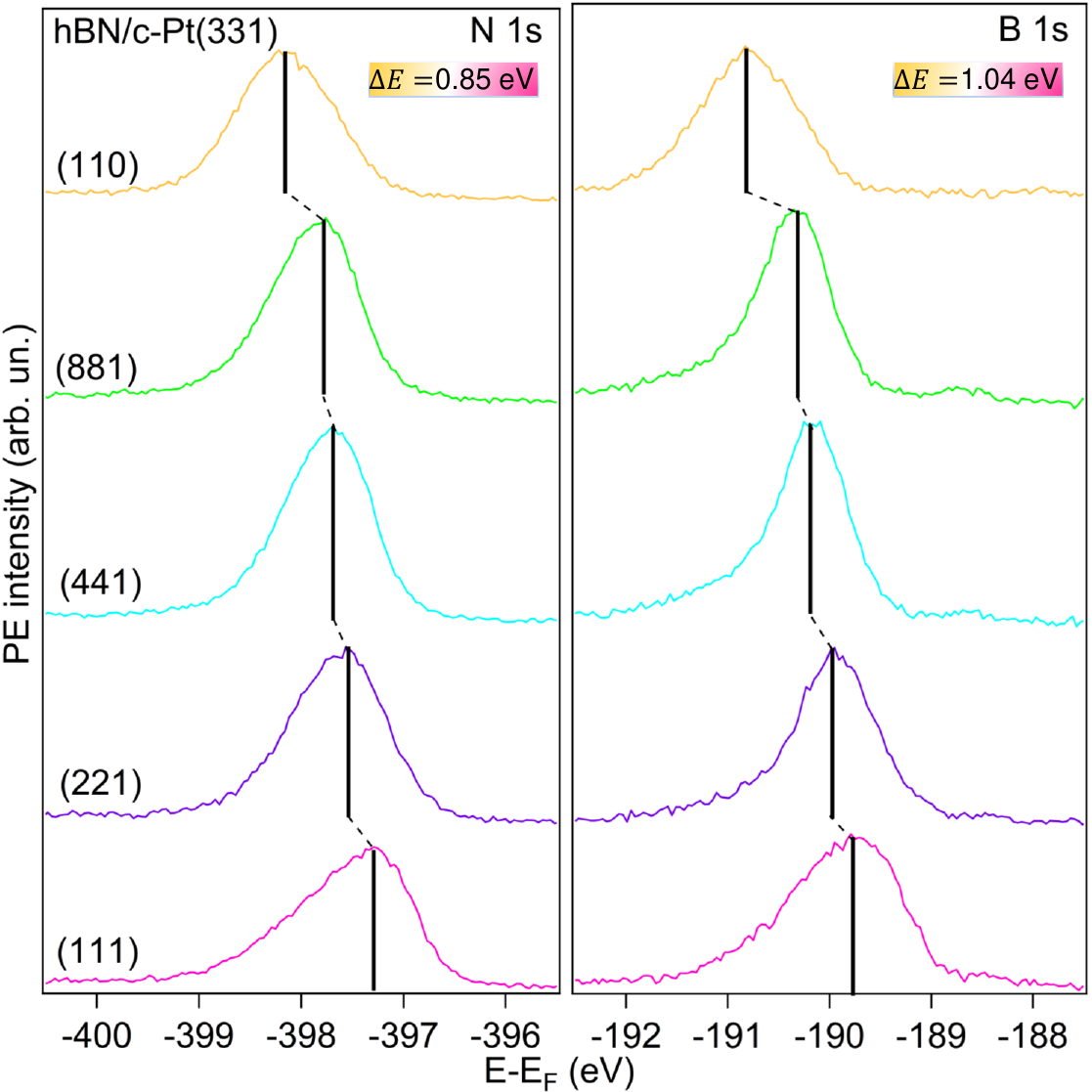}
    \caption{XPS analysis of hBN grown on c-Pt(331) substrate. The N and B 1s XPS core levels spectra were taken at different positions on the curved crystal with a photon energy of 480 eV.}
    \label{fig:4}
\end{figure}

\begin{table}[ht!]
\begin{center}
\scalebox{0.65}{
\begin{tabular}{|c|c|c|c|c|c|c|} 
\hline
Stable facet&  & Pt(110) & Pt(881)& Pt(441)& Pt(221)& Pt(111) \\
\hline
Peak position (eV) &N 1s  &-398.14 &-397.76 &-397.68 &-397.56 &-397.29   \\
&B 1s  &-190.83 &-190.31 &-190.17 &-189.99 &-189.79   \\
\hline
\end{tabular}}
\caption{Peak positions of N and B 1s XPS core level spectra taken after the growth of hBN on c-Pt(331). These spectra were captured at the stable facets of (110), (881), (441), (221), and (111).}
\label{Tab:1}
\end{center}
\end{table}

\begin{figure*}[ht!]
    \centering
    \includegraphics[scale=0.5]{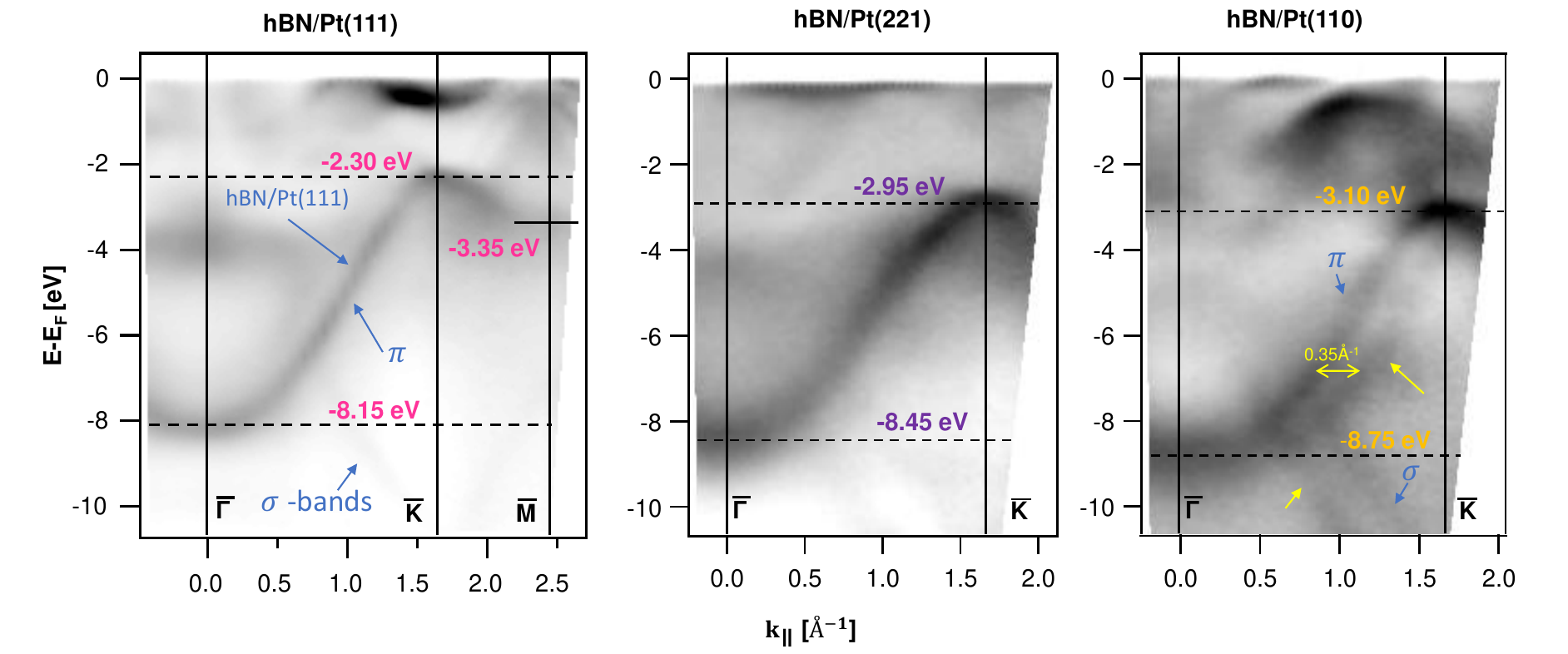}
    \caption{\textbf{ARPES analysis} photoemission intensity maps along the $\bar{\Gamma}$-$\bar{\text{K}}$-$\bar{\text{M}}$ direction of the surface BZ of the hBN on c-Pt(331) at the (111), (221) and (110) facets, measured with $h\nu$ = 72 eV. The $\pi$- and $\sigma$-bands are indicated with blue arrows; in the hBN/Pt(110) system, yellow arrows point to the $\pi$ band replica; the separation between them is $0.35~\mathring{\mathrm{A}}$.}
    \label{fig:5}
\end{figure*}
Following the hBN growth, the electronic and chemical properties were characterized by means of XPS, as shown in Figure \ref{fig:4}. Core-level photoemission spectra of N and B 1s were acquired at different positions along the curvature. In particular, at the previously observed stable facet positions. The B and N 1s spectra display a systematic shift toward higher binding energy (leftward shift) when moving from the (111) to the (110) facet, with the peak positions presented in Tab. \ref{Tab:1}. Hence, at Pt(111), the  N and B 1s peak positions are consistent with reported studies~\cite{nappini2016formation, preobrajenski2007influence}. This shift signifies an increase in interaction strength, highlighting a distinct trend from the hBN/c-Ni(111) system~\cite{fernandez2019boron}, where the interaction strength was highest at the (111) facet. The N and B 1s spectra of hBN on Pt facets shift by $\Delta E$ of around 0.85 eV and 1.04 eV, respectively, moving from the (111) to the (110) facet. In addition, the NEXAFS analysis presented in Fig. S2 in SM~\cite{SM} for hBN on a c-Pt(331) substrate is consistent with the XPS findings. This trend confirms that substrate orientation is crucial in modifying overlayer/substrate interactions.

To further study the electronic properties of hBN/c-Pt(331), the ARPES technique was employed. Here, the electronic band structure of hBN grown on curved Pt(331) was probed by ARPES ($h\nu$ = 72 eV). Noteworthy that the valence band maximum (VBM) of the hBN monolayer serves as a reliable indicator of the strength of its chemical interaction with the underlying substrate. Thus, we configured the ARPES setup in a way that reaches the $\bar{\text{K}}$ point of hBN, its VBM, by using angular scans, later converted into the $\text{k}_{\parallel}$ in all the following measurements. The normal emission ($\theta = 0$, tilt angle $\phi = 0$) is aligned along the [11$\bar{2}$] direction of the bulk of the crystal.
Photoemission intensity maps along $\bar{\Gamma} \bar{\text{K}}$ were recorded for the hBN/c-Pt(331) system at the positions of the stable facets (111), (221), and (110), shown in Fig. \ref{fig:5}. For the hBN/Pt(111) system, the $\pi$ band minimum was observed at the $\bar{\Gamma}$ point at -8.15 eV, while the maximum at the $\bar{\text{K}}$ point was around -2.3 eV. Additionally, the $\bar{\text{M}}$ point position was measured at -3.35 eV. Similar measurements were conducted for the hBN/Pt(221) and hBN/Pt(110) systems; a downward shift of the $\pi$ band is observed as one moves away from the (111) facet. The $\sigma$ band maximum is visible around -4.00 eV for hBN/Pt(111), shifting to approx. -5.00 eV for hBN/Pt(110), though it appears blurry due to the excitation energy used. Furthermore, in the hBN/Pt(110) system, $\pi$- and $\sigma$-band replica features in Fig. \ref{fig:5} are observed, indicated by the yellow arrows. These features are attributed to superstructure-induced Umklapp replicas of the hBN bands, in agreement with the findings of Thaler et al.~\cite{thaler2020single}, and similarly observed for hBN/Rh(337) by Ali et al.~\cite{ali2021atomically}. The hBN/Pt(110) exhibits a superstructure of $(1\times n)$ missing-row reconstruction of Pt(110), with $n=5$ or $6$ ~\cite{thaler2020single}. Due to this superstructure, additional reciprocal-lattice vectors emerge; therefore, for a $\pi$ band with dispersion $E_{\pi}(\mathbf{k})$, the replica is expected at $E_{\pi}(\mathbf{k}\pm\mathbf{G}_{\mathrm{S}})$. The momentum separation between the main $\pi$ band and its replica, indicated by a yellow double-sided arrow, is $\Delta k=0.35~\mathring{\mathrm{A}}^{-1}$ along the [1$\bar{1}$0] direction. The $\Delta k$ corresponds to the magnitude of the superstructure reciprocal-lattice vector, $|\mathbf{G}_{\mathrm{S}}|$, for a first-order Umklapp process. Subsequently, the real-space periodicity is $L_{\mathrm{S}}=2\pi/\Delta k\approx17.9~\mathring{\mathrm{A}}$. Now, we can compare using the Pt lattice constant $a_{\mathrm{Pt}}=3.92~\mathring{\mathrm{A}}$~\cite{ashcroft1976solid}, the surface lattice spacing along [1$\bar{1}$0] is $d_{[1\bar{1}0]}=a_{\mathrm{Pt}}/\sqrt{2}=2.77~\mathring{\mathrm{A}}$. A $(1\times6)$ periodicity consequently gives $L_{1\times6}=6d_{[1\bar{1}0]}=16.6~\mathring{\mathrm{A}}$ and $G_{1\times6}=2\pi/L_{1\times6}\approx0.38~\mathrm{\mathring{\mathrm{A}}^{-1}}$. The close correspondence between the measured $\Delta k\approx0.35~\mathring{\mathrm{A}}$ and the expected $G_{1\times6}\approx0.38~\mathring{\mathrm{A}}$ therefore supports our interpretation of the observed $\pi$ band replica as superstructure-induced Umklapp scattering. 

\section{Density Functional Theory Results}\label{sec:4} 
\begin{figure*}[ht!]
    \centering
    \includegraphics[scale=0.82]{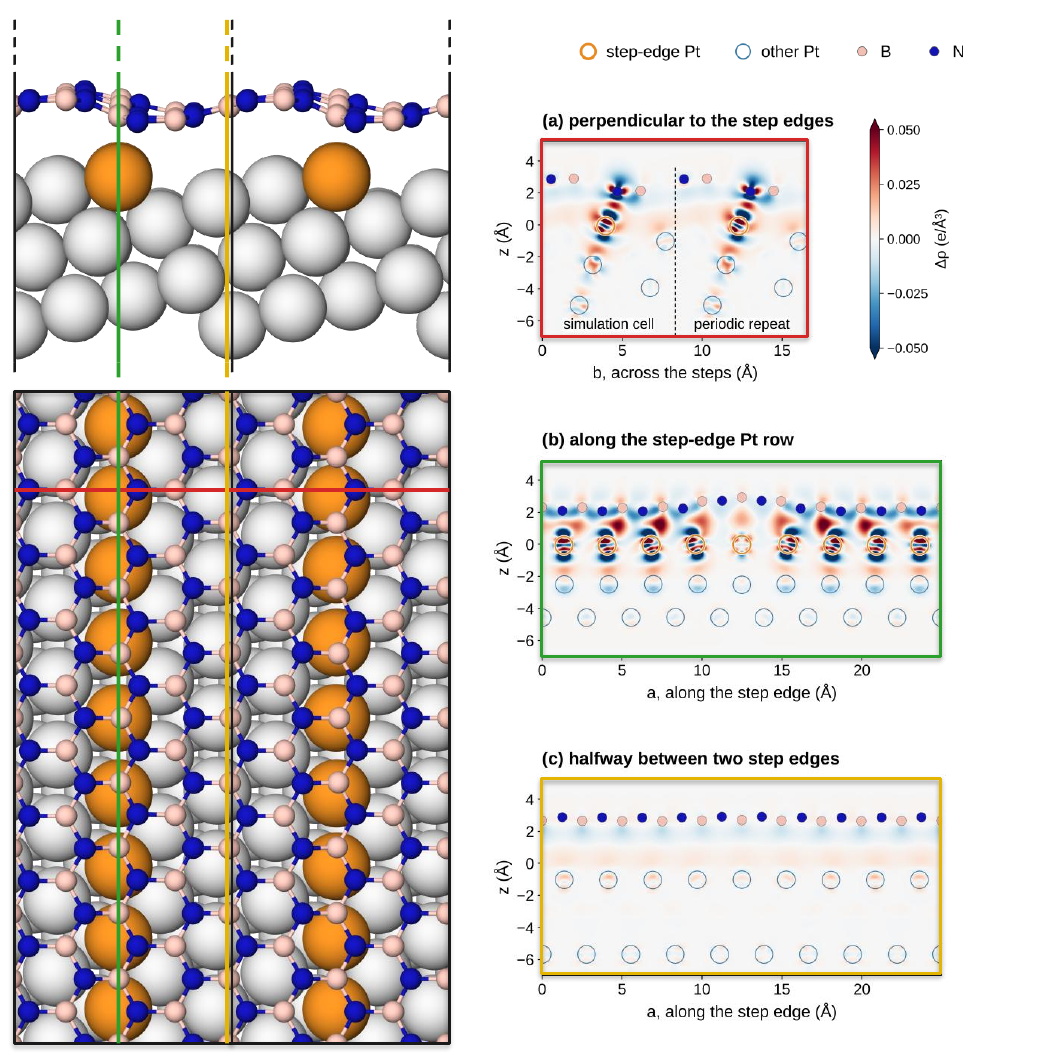}
    \caption{Left panels: side and top view of the DFT-optimized structural model for the hBN/Pt(221) configuration. Black lines indicate the periodically repeated simulation cell (in the side view only a part of the vacuum region is represented). Right panels: electron density rearrangement $\Delta\rho$ due to the overlayer-substrate interaction (difference between the electron density of the entire system and of the two constituents). Plots are on planes indicated by the corresponding coloured lines in the model. On the plots, the projected positions of the atoms closest to the plane are indicated.}
    \label{fig:6}
\end{figure*}

To rationalize the experimental findings, we investigate by DFT  the hBN overlayer on experimentally observed stable and unstable facets. Even for hBN on the flat Pt(111) surface, there is a strong lattice mismatch with the substrate.
The calculated equilibrium lattice parameters of bulk Pt and hBN are respectively ${a}^{Pt}$=3.92 {\AA} and ${a}^{hBN}$=2.515 {\AA}, very close to the experimental values~\cite{ashcroft1976solid,pease1952x}, allowing us to simulate our system on a realistic ground. The Pt interatomic distance on the (111) terraces and along the [$\bar{1}$10] step edges is thus ${a}_{(111)}^{Pt}={a}_{\parallel}^{Pt}$=2.77 \AA; therefore, hBN has a mismatch $(a^{hBN}-{a}_{(111)}^{Pt})/{a}_{(111)}^{Pt}=-9.2\%$  with respect to Pt(111).
For a realistic description, in our simulations we keep the substrate unstrained; therefore, we refer to the {\em{strain}} that has to be applied to hBN to accommodate it on the substrate, i.e.  $(a^{hBN~strained}- a^{hBN})/{a}^{hBN}$.

Before discussing the results for hBN on the vicinals of Pt(111), it is instructive to review the main results for hBN/Pt(111), extensively reported in SM~\cite{SM}. Our tests on highly strained (1 $\times$ 1) hBN on (1 $\times$ 1) Pt indicate that configurations with N on top of Pt and B in hollow sites are slightly more stable than others, although it must be noted that in mismatched interfaces the relative registry changes continuously. A good matching between hBN and Pt(111) is achieved with a (10 $\times$ 10) hBN supercell epitaxially aligned on  (9 $\times$ 9) Pt~\cite{preobrajenski2007influence}, where hBN shows a compressive strain of only $-$0.9\% (left panels of Fig. S3 and upper part of Table 1 in SM~\cite{SM}). 

Our DFT results show that the epitaxially aligned domains (R0)  are favoured with respect to 30$^{\circ}$-oriented domains (R30), where the armchair direction of hBN is aligned to the close-packed $[\bar{1}10]$ direction (right panels of Fig. S3 and lower part of Table 1 in SM~\cite{SM}). This is consistent with the experimentally observed alignment of the zig-zag direction of hBN along the step edges of the vicinals of Pt(111).
\begin{figure*}[hbt!]
    \centering
    \includegraphics[scale=0.52]{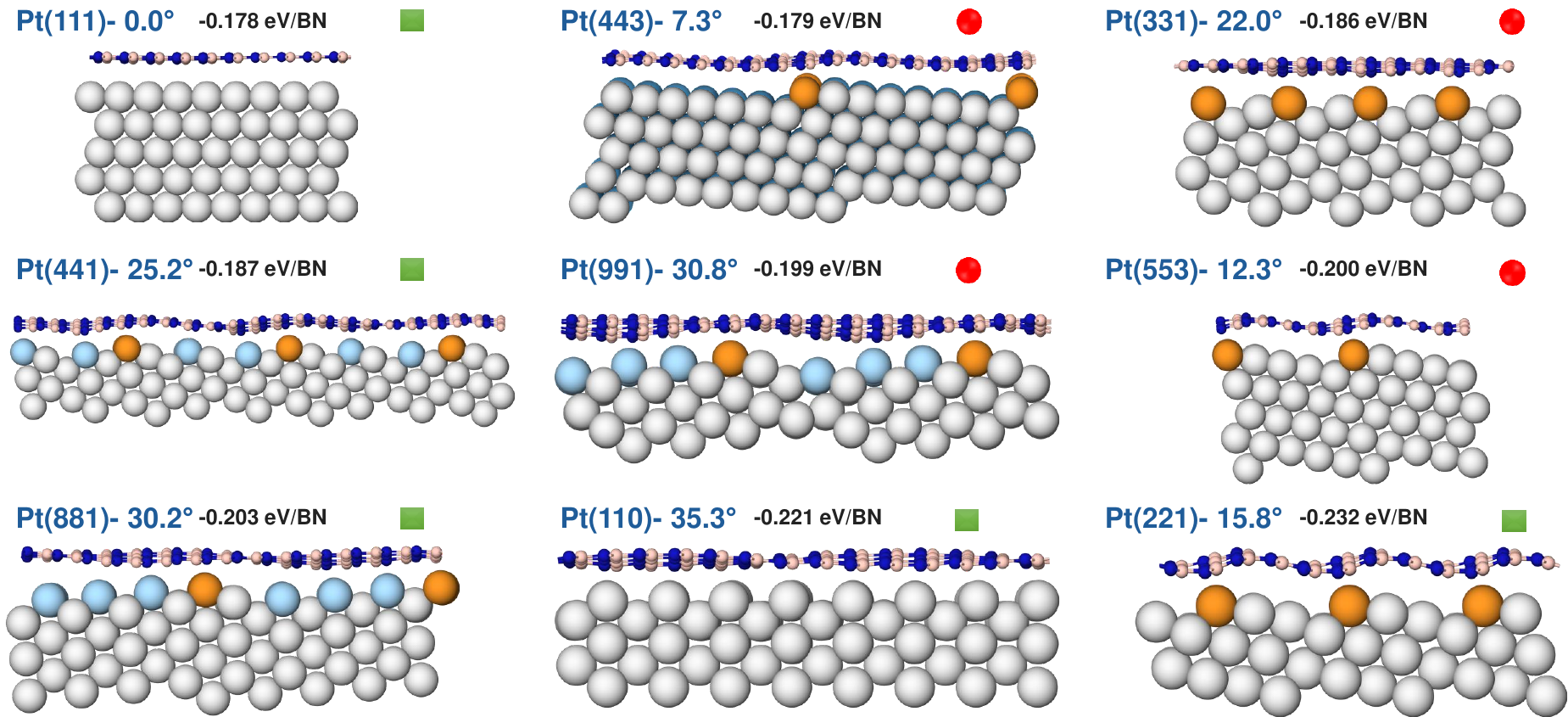}
    \caption{Side view of the DFT-optimized models of the most representative structures for Pt(111) and various vicinals. Step edges are highlighted in orange, and (110)-type surface atoms in light blue. Red dots and green squares identify the experimentally unstable/stable facets after hBN growth. The DFT adhesion energy per BN pair is reported.}
    \label{fig:7}
\end{figure*}

\begin{figure}[hbt!]
    \centering
    \includegraphics[scale=0.31]{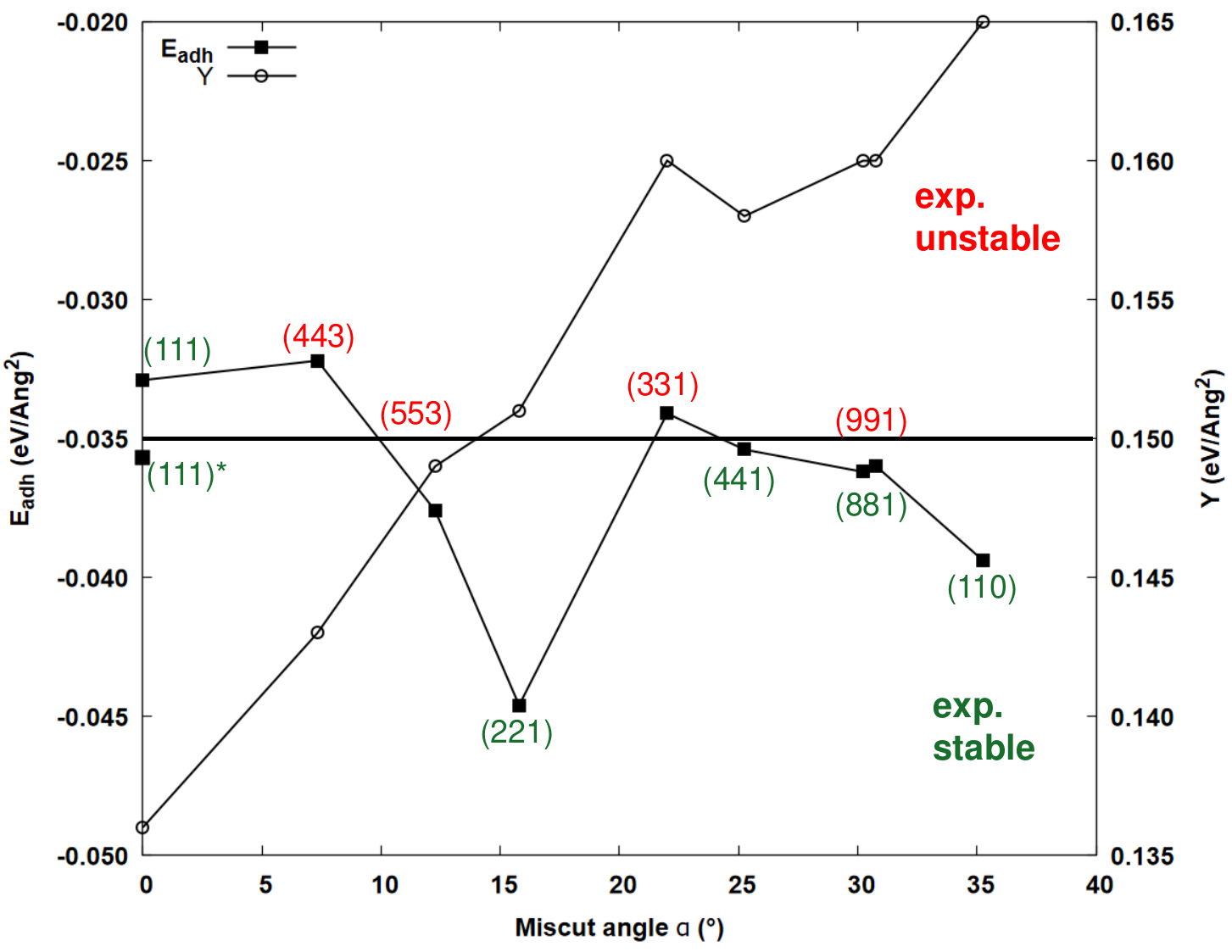}
    \caption{Adhesion energy of hBN on Pt(111) and on various vicinals plotted as a function of the miscut angle and free energy Y of the corresponding bare surface, both in eV/{\AA}$^2$.  The range of the two scales is the same to facilitate the comparison between the plotted quantities. For a meaningful comparison, lines join models with similar strain on hBN; the (111)* labelled result refers to a cell with higher strain.  The horizontal thin line is in the middle of the free energy range and is only a guide to the eye. Green(red) labels distinguish experimentally observed stable(unstable) facets.}
    \label{fig:8}
\end{figure} 
As mentioned earlier, hBN was widely studied theoretically on flat TM substrates but only scarcely studied on vicinal surfaces~\cite{Grad_2003, laskowski2010ab, Arias_2018, farwick_2016}. 
In fact, such an investigation is particularly challenging for extended domains. Firstly, it requires the construction of supercells that match the periodicity of hBN to that of the Pt facets. The in-plane periodicity of the latter depends on the miscut angle and can be quite large for certain facets Table 2 of SM~\cite{SM} and Figs. S6-S7). The additional constraint of aligning the zig-zag direction of hBN with the Pt step edges makes the construction of commensurate supercells even more challenging.  Taking into account such a constraint, we search for quasi-commensurate multiples of the Pt surface cells (rectangular or oblique, depending on the miscut angle $\alpha$) and of the hBN \emph{rectangular} unit cell with the short(long) side equal to the zig-zag (armchair) periodicity Fig. S5 in SM~\cite{SM}). 

Suitable structural models must be a compromise between a realistic description with a small strain applied to hBN and an affordable size. 
Aligning the short side of 10 rectangular hBN cells with 9 Pt atoms along the step edges, the strain on hBN along its zig-zag direction is  $-$0.9\%. Different facets require different geometries and strain along the armchair direction. 
The simplest example is shown in Fig.~\ref{fig:6} for hBN/Pt(221). The periodically repeated simulation cell is the smallest (276 atoms) among those considered for the various facets. This is due to the short step-to-step distance (8.36 {\AA}), the orthogonality of the vectors defining the bare surface unit cell, and the favorable matching between the hBN periodicity along the armchair direction and the step spacing, which can be accommodated with a moderate compressive strain on hBN (4.5\%) only with two repetitions of its unit cells. The right panels of Fig.~\ref{fig:6} show the electron density rearrangement due to the overlayer-substrate interaction (difference between the electron density of the entire system and of the two constituents) plotted on different planes. The presence of almost equivalent blue/red regions indicates that the charge transfer between the overlayer and substrate is negligible or very small. A huge electron rearrangement occurs at the step edges, suggesting a stronger binding with respect to the terraces.

Table 3 in SM~\cite{SM} summarizes the structural features of models for hBN with strain smaller than 5\% covering different facets. The best DFT-optimized models are shown in Fig. \ref{fig:7}.

To compare the stability of vicinals under the hBN overlayer, some insight comes from the {\em{adhesion energy}} E$_{\text{adh}}$ per BN pair, calculated by subtracting to the total energy of the combined hBN/Pt system ($E_{\text{hBN/Pt}}$) from those of the corresponding isolated hBN monolayer ($E_{\text{hBN}}$) and Pt slab ($E_{\text{Pt slab}}$), each one separately optimized in the same supercell of the combined system, then dividing by the number of BN pairs $\text{N}_{BN}$:
\begin{equation}
\label{Eq:Eads}
E_{\text{adh}} = \left (E_{\text{hBN/Pt}} - E_{\text{Pt slab}} - E_{\text{hBN}}\right )/\text{N}_{BN}.
\end{equation}
$E_{adh}$  (reported in Table 1 and Table 3 of SM~\cite{SM} for hBN on (111) surface and its vicinals, respectively) varies from $-$0.17 eV to $-$0.23 eV. A numerical uncertainty of 0.01 eV due to technical parameters such as kinetic energy cutoff and Brillouin zone sampling should be attributed to these values. An additional and similar uncertainty also comes from the strain state of hBN, as shown by comparing the values obtained with different models for (441) and (881). Although a variation of $E_{adh}$ on different vicinals can be appreciated, we cannot distinguish the same trend observed in experiments: (221) and (111) show the strongest adhesion energy, but (441) and (881) compete at least with (331) and (991). It is important to notice that the adhesion energy provides information on the relative preference of the overlayer for covering one vicinal surface over another, but it does not measure the stabilization of a given vicinal surface with respect to its corresponding clean surface and, more generally, does not determine the relative stability among different vicinals. Therefore, the DFT results on $E_{adh}$ do not indicate a discrepancy with experiments, but simply that the average strength of overlayer-surface interaction is not the only ingredient of the selective faceting mechanism.
Further insight is given by the  {\em{ clean-surface free energy}}, which in the simplest approximation, neglecting the vibrational entropy, is given by:
\begin{equation}
\label{Eq:gammaclean}
\gamma=
\left (E_{\text{Pt slab}}-{\text N}_{Pt}\mu_{Pt}\right )/{2A}
\end{equation}

where ${\text N}_{Pt}$ are the number of Pt atoms of the slab, $\mu_{Pt}$ the Pt chemical potential, $A$ the surface area for each side of the slab. DFT-calculated values of $\gamma$ are reported in Table 2 of SM~\cite{SM}.
The estimate of $\gamma$ and $E_{adh}$ is not sufficient to clearly disentangle overlayer-driven and substrate-driven contributions to vicinal selection upon hBN covering, but their comparison, shown in Fig.~\ref{fig:8}, gives some indications. The two quantities are shown using the same units and on the same scale to facilitate comparison. The almost monotonic behaviour of $\gamma$ with the miscut angle $\alpha$, with an overall variation of 0.03 eV/{\AA}$^2$, suggests an increasing instability of clean vicinals. On the other hand, $E_{adh}$ is weaker on some surfaces offering a low density of step edges, such as (443) and (553) and in the limiting case of the flat (111) surface, suggesting that localized covalent bonding at step edges establishes a stronger interaction with the substrate.
On the basis of DFT and experimental results, we can conclude that hBN growth changes the relative stability of Pt vicinals and several factors are involved: in some cases, such as (110) and (221),  a particularly strong interaction of hBN stabilizes these surfaces that would otherwise be intrinsically less stable than others, whereas the observed stability of the covered (111) could be ascribed to the high stability of the clean surface.

\section{Conclusion}
In summary, the structural and electronic properties of hBN grown on a curved Pt(331) crystal were investigated employing various techniques such as LEED, STM, XPS and ARPES, clearly detecting five stable facets Pt(111), Pt(221), Pt(441), Pt(881), and Pt(110) with an increased interaction strength from (111) to (110). Accurate DFT numerical simulations revealed that the stabilization of facets upon hBN growth on c-Pt(331) is influenced by multiple factors that cannot be fully disentangled and whose relative importance varies from one facet to another: in some cases the clean substrate energetics dominates, in others the overlayer-substrate interaction. In general, step edges play a relevant role in binding the overlayer. Furthermore, the alignment of the zig-zag direction of hBN with them and the $\langle 110\rangle$ directions of Pt is preferred with respect to the armchair direction. To conclude, by performing static DFT calculations, we provided a thermodynamic explanation for the facet morphology, but not a kinetic pathway of the faceting mechanism or vicinal-surface evolution under the growing overlayer.
In general, the observed faceting is a complex energy-minimization process involving not only the chemical coupling between the overlayer and the substrate but also surface stress, step energy, step--step interactions, varying across the curved crystal and upon hBN covering.

\section{Declaration of competing interests}

There are no conflicts to declare.

\section{Data availability}
Data will be available upon request from the authors.

\section{Acknowledgments}

We acknowledge financial support from grants PID2023-149158OB-C44 funded by the Spanish MCIN/AEI/10.13039/501100011033 and the Basque Government (Grant IT2133-26). Part of the research leading to the results has been supported by the project CALIPSOplus under Grant Agreement 730872 under the EU Framework Programme for Research and Innovation HORIZON 2020.  I. P. and F. B. acknowledge financial support from EUROFEL project (RoadMap Esfri). Additionally, A. M. I. B. express gratitude to Fomento San Sebastián and COST-action OPERA for their funding contributions. M.P. and S.M.M. acknowledge financial support from the NextGenerationEU Programme (CUP Grant. No. J92B24001090007, PNRR Investimento M4.C1.3.4.T1 and M4.C1.4.1 -- DM 629/2024) and the CINECA award under the ISCRA initiative, for the availability of high-performance computing resources and support. We also acknowledge the Helmholtz-Zentrum Berlin für Materialien und Energie Facility for the allocation of beam time on the GELEM Dipole.

\appendix
\section{Supplemental Material}
\label{app1}

\bibliography{cas-refs}
\bibliographystyle{elsarticle-num}

\end{document}


\begin{frontmatter}

\title{Supplemental Material\\ 
{\bf Selective faceting induced by Hexagonal Boron Nitride on curved Platinum substrate}}

\author[1,2,3]{Alaa Mohammed Idris Bakhit\corref{cor1}}
\author[4]{Muhammad Saad Mahmood}
\author[5,6]{Khadiza Ali}
\author[7]{Anna A. Makarova}
\author[8,9]{Igor Piš}
\author[9]{Federica Bondino}
\author[1]{Frederik Schiller}
\author[4]{Maria Peressi\corref{cor2}}

\address[1]{Centro de F\'{\i}sica de Materiales (CSIC-UPV-EHU) and Materials Physics Center (MPC), E-20018 San Sebasti\'an, Spain}
\address[2]{Departamento de F\'{\i}sica Aplicada I, Universidad del Pa\'{\i}s Vasco UPV/EHU, 20018 San Sebasti\'an, Spain}
\address[3]{Donostia International Physics Center, 20018 Donostia-San Sebasti\'an, Spain.}
\address[4]{Department of Physics,  University of Trieste,  Via Alfonso Valerio, 2, 34127 Trieste, Italy.}
\address[5]{Department of Physics, BITS Pilani, Hyderabad Campus, Hyderabad, 500078, Telangana, India}
\address[6]{Chalmers University of Technology, Göteborg, Chalmersplatsen 4, 412 96 Göteborg, Sweden}
\address[7]{Physikalische Chemie, Institut für Chemie und Biochemie, Freie Universität Berlin, Arnimallee 22, 14195 Berlin, Germany}
\address[8]{Institute of Electrical Engineering, Slovak Academy of Sciences, Dúbravská cesta 9, 84104 Bratislava, Slovakia}
\address[9]{CNR - Istituto Officina dei Materiali (IOM), S.S. 14 km 163.5, 34149 Basovizza, Trieste, Italy}

\cortext[cor1]{alaa.mohammed$@$dipc.org}
\cortext[cor2]{peressi$@$units.it}

\end{frontmatter}

\section*{LEED analysis}

\begin{figure*}[ht!]
    \centering
    \includegraphics[scale=0.4]{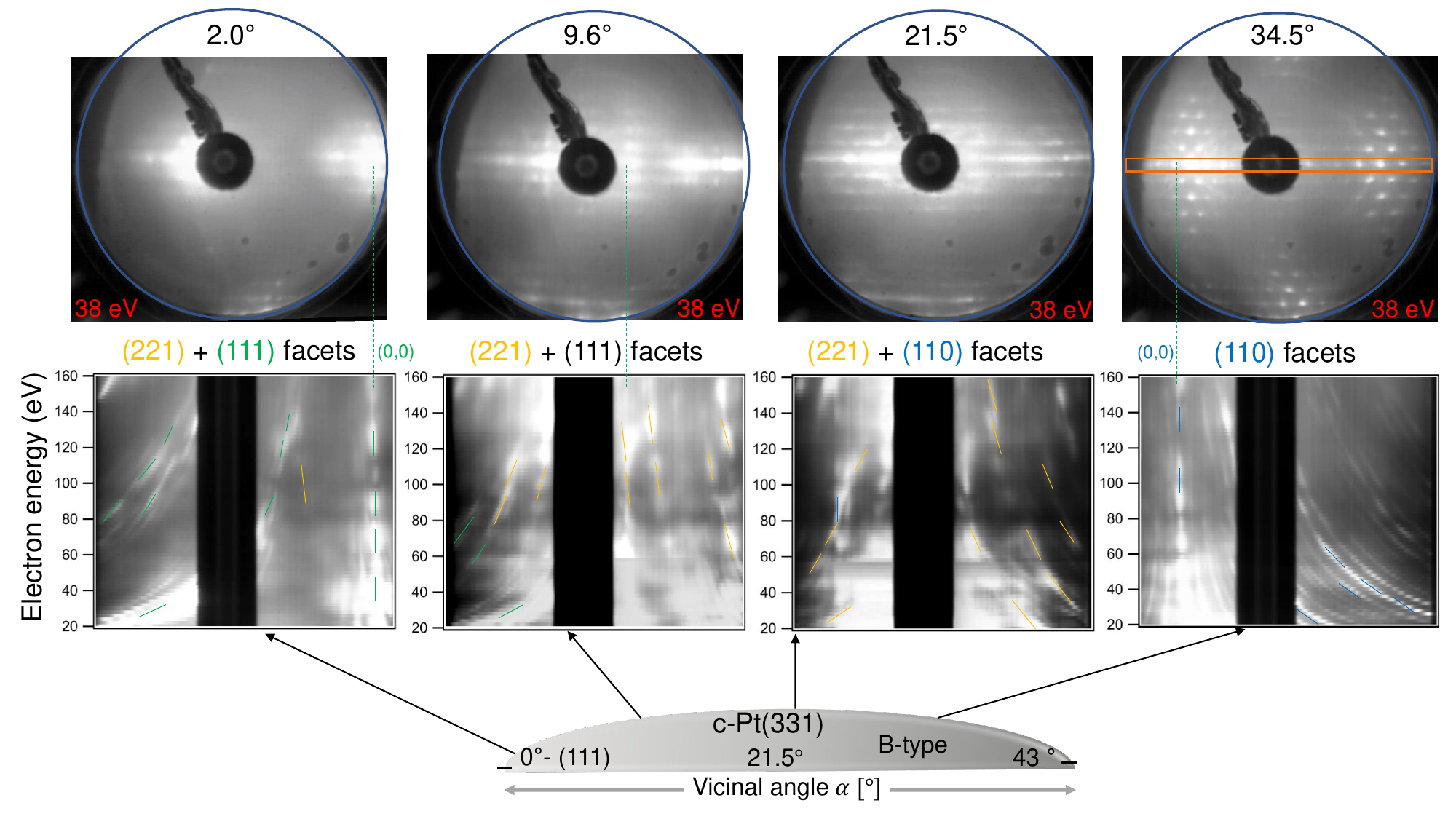}
    \caption{LEED IV Analysis of hBN grown on curved Pt(331): top panel LEED pattern at the indicated positions taken at kinetic energy of 38 eV. The bottom panel LEED IV image profile was taken at the central line indicated by the orange rectangle. The green lines denote the (0,0) spots of the stable facets indicated at the top of the LEED IV images. The green (111), yellow (221), and blue (110) lines point towards the stable facets' (0,0) spot~\cite{bakhit2022structure}.}
    \label{fig:S1}
\end{figure*}
The stable facets formed in the case of the hBN/c-Pt(331) system were confirmed by the results obtained from the LEED IV analysis of the diffraction spots, shown in Fig. \ref{fig:S1}. In the top panel, one can see the LEED images at different positions (vicinal angles) of the curved crystal by lateral sample movement. The line in green indicates that the only electron energy-independent spot at each position is the (0,0) spot, indicating the stable facet(s). These stable facets are easily observed in the energy-dispersive IV mapping in the bottom panel. Our observations directly point to the following stable facets: (111), (221), and (110).

\section*{NEXAFS analysis}

\begin{figure}[ht!]
    \centering
    \includegraphics[scale=0.6]{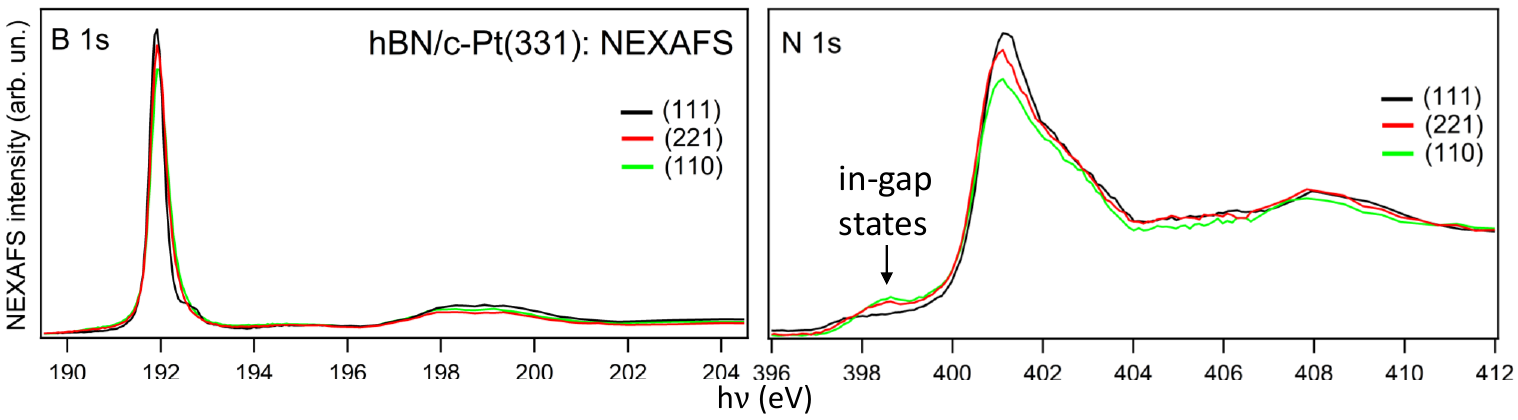}
   \caption{NEXAFS analysis of hBN grown on c-Pt(331) substrate. NEXAFS at the B and N $K$ adsorption edges acquired at the stable facets, Pt(111), Pt(221), and Pt(110)~\cite{bakhit2022structure}.}
   \label{fig:S2}
\end{figure}

Fig. \ref{fig:S2} presents the B and N $K$-edge X-ray absorption spectra recorded at the (111), (221), and (110) facets. The weakest interaction is observed for the hBN/Pt(111) system, with the interaction strength increasing as one moves away from the (111) surface. Furthermore, the in-gap states are more pronounced in the N $K$-edge spectra on the (221) and (110) facets, while the sharpest peak appears on the (111) facet. This trend provides clear evidence of the variation in the interaction strength across different crystal orientations. Overall, vicinal surfaces show strong interaction. Again, NEXAFS analysis further validates the importance of substrate orientation in the overlayer-substrate interaction strength variation; e.g., in the hBN/c-Ni(111) system, hBN/Ni(111) is the strongest interacting system~\cite{fernandez2019boron}. In contrast, hBN/Pt(111) is the weakest interacting system~\cite{bakhit2022structure}.

\section*{Numerical simulations of hBN on Pt(111)}

As reported in the Main Text, there is a high lattice mismatch between hBN and Pt(111) (${\bf a}^{hBN}$= 2.525 {\AA}, ${\bf a}_{(111)}^{Pt}$=${\bf a}_{bulk}^{Pt}$/$\sqrt{2}$=3.92 \AA/$\sqrt{2}$= 2.77 {\AA}). Reliable models to describe hBN on Pt(111) require common simulation cells made by almost commensurate supercells of Pt and hBN, with a small strain on hBN to match the underlying substrate, which is kept unstrained.

\subsection*{Epitaxially aligned geometry and strain effects}

\begin{figure*}[htb!]
    \centering
    \includegraphics[scale=0.6]{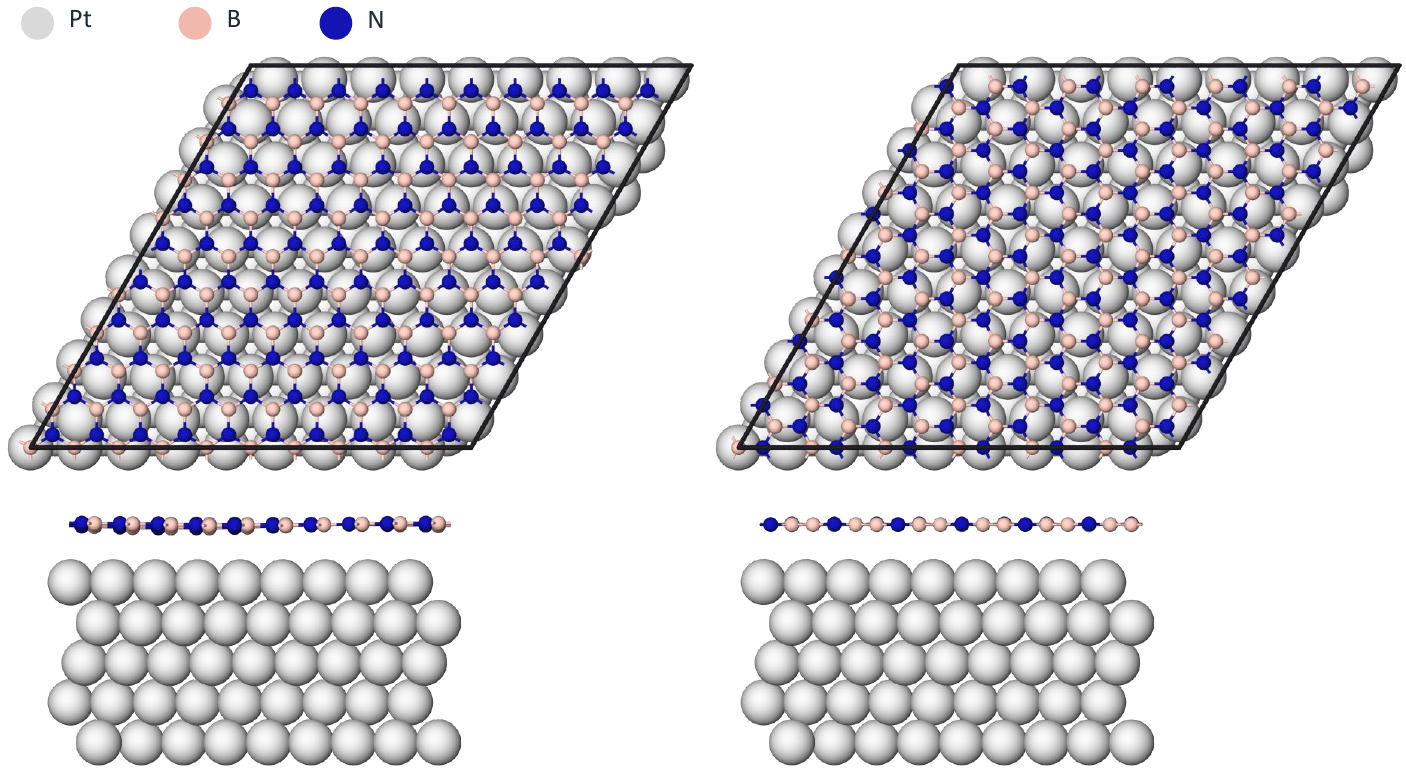}
    \caption{Top and side view of the DFT-optimized supercell model for hBN/Pt(111) in R0 (left panels) and R30 (right panels) configurations. For R0 the supercell contains (10$\times$10)hBN/((9$\times$9)Pt. For R30 the supercell contains (6$\times$6)hBN$_{hex~armchair}$/((9$\times$9)Pt, i.e. a 3$\times$3 multiple of the almost commensurate  (2$\times$2)hBN$_{hex~armchair}$/((3$\times$3)Pt cell. }
    \label{fig:S3}
\end{figure*}

We first consider hBN epitaxially aligned on Pt, i.e., the zig-zag directions of hBN aligned with the $\langle 110\rangle$ directions of the Pt surface: we label this orientation R0, corresponding to a  {\em{rotation}} angle of zero between the two directions. The supercells are hexagonal since the overlayer and substrate share the same two-dimensional Bravais lattice symmetry. Tab.~\ref{Tab:1} summarizes the structural details of possible almost commensurate supercells with a number of BN pairs not exceeding 150.
The supercell with (10 $\times$ 10) hBN unit cells  aligned on  (9 $\times$ 9) Pt unit cells (with  $-$0.9\% strain on hBN)
(Fig. \ref{fig:S3}) has already been proposed in the literature.  Our results give an almost flat overlayer with minimal corrugation (0.29 \AA), an interfacial distance with an average value of 3.41 {\AA} and a minimum value of 3.11 {\AA}, and an adhesion energy (see Main Text for definition) $E_{adh}=-$0.18 eV/BN pair. Our results indicate that hBN binds to Pt with a strength similar to that of graphene on Ni,  weaker if compared to other more strongly interacting systems such as hBN on Rh(111) and Ru(001)~\cite{laskowski2010ab,preobrajenski2007influence}. Our findings are partially comparable with those of Ref. ~\cite{laskowski2010ab}, which does not include dispersion forces, reporting an average overlayer-substrate distance around 2.7 {\AA}, a corrugation of about 0.5 {\AA}, and a smaller adhesion energy,  $E_{adh}=-$0.07 eV/BN pair.  This result, from literature and our additional tests performed without including dispersion corrections, gives much smaller adhesion energies, indicating that van der Waals interactions play a relevant role in the interfacial bonding.

\begin{table*}[htb!]
\begin{center}
\scalebox{0.9}{\begin{tabular}{|c|c|c|c|c|}
\hline
no. hBN unit cells & no. Pt unit cells & strain on hBN (\%) & no. BN pairs  & $E_{adh}$ \\
\hline
\hline
\multicolumn{5}{|l|}{R0}\\ 
\hline
(12 $\times$ 12) & (11 $\times$ 11)  & +1.0& 144 & \\
(11 $\times$ 11)& (10  $\times$ 10) & +0.1& 121 & \\
(10 $\times$ 10)& (9  $\times$ 9) & $-$0.9 & 100 & $-$0.18\\
(9 $\times$ 9) & (8 $\times$  8) & $-$2.1 & 81 &\\
(8 $\times$ 8) & (7 $\times$  7)  & $-$3.6 & 64 & $-$0.21\\
\hline
\hline
\multicolumn{5}{|l|}{R30}\\ 
\hline
(2 $\times$ 2) (or (6  $\times$ 6)) & (3  $\times$ 3) (or (9  $\times$ 9)) & $-$4.7 & 12 (108) & $-$0.17\\
(5  $\times$  5)& (8  $\times$ 8)  & 1.7& 75 & \\
(7  $\times$ 7) & (11  $\times$ 11)& $-$0.2 & 147 & \\
\hline
\end{tabular}}
\caption{Approximately commensurate supercells for hBN on Pt(111) with less than 150 BN pairs. Upper panel: hBN epitaxially aligned on Pt(111), with the hBN zig-zag directions aligned along the Pt $\langle 110\rangle$ directions (R0 domain). The reference unit cells of hBN and Pt are the primitive hexagonal unit cells with lattice parameter of 2.515 {\AA} and 2.77 {\AA} respectively. Lower panel: configurations with the hBN zig-zag directions rotated by 30$^\circ$ with respect to the Pt $\langle 110\rangle$ directions, i.e., armchair hBN directions aligned along the Pt $\langle 110\rangle$ directions (R30 domain). The reference unit cells of hBN and Pt are both hexagonal with lattice parameter of 4.36 {\AA} and 2.77 {\AA} respectively. The (2$\times$ 2)hBN$_{armchair}$/(3$\times$ 3)Pt is equivalent to the (6$\times$ 6)hBN$_{armchair}$/(9$\times$ 9)Pt, which has been actually used with $\Gamma$ point only for Brillouin zone integration.   From left to right: number of hBN and of Pt unit cells in both primitive lattice directions; strain applied to hBN to match the Pt supercell; number of BN pairs; adhesion energy per BN pair. See Fig.~\protect{\ref{fig:S5}} for a sketch of the hBN unit cells.}
\label{Tab:1}
\end{center}
\end{table*}

\begin{figure*}[htb!]
    \centering
    \includegraphics[scale=0.35]{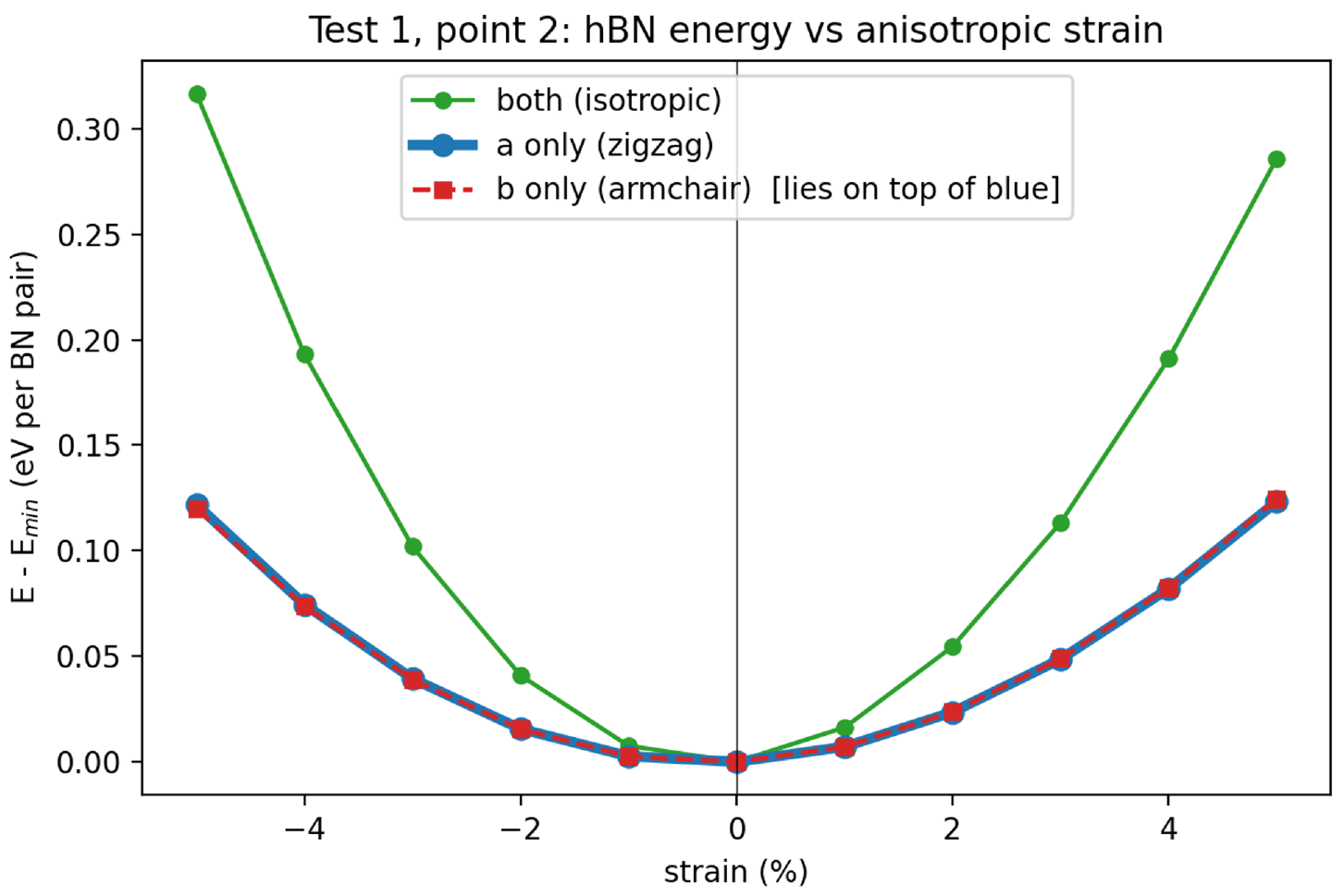}
    \caption{Total energy variation of hBN under compressive and tensile strain. Besides the isotropic strain, the effect of uniaxial strain applied to the zig-zag or the armchair directions.}
    \label{Fig:S4}
\end{figure*}

Tab.~\ref{Tab:1} also reports our results for an alternative supercell with higher compressive strain, (8 $\times$ 8)hBN/(7 $\times$ 7)Pt, where $E_{adh}=-$0.21 eV. 
It is instructive to discuss the origin of this variation. A partial cancellation of the strain contribution to the total energies of the supported and free-standing overlayer is expected in the calculated adhesion energy. Indeed, the total energy per BN pair increases by 0.15 eV in the free-standing case when the isotropic strain changes from $-$0.9\% to $-$3.6\% (Fig.~\ref{Fig:S4}), whereas the corresponding variation in $E_{adh}$ for the supported system is only $-$0.03 eV. We can conclude that $E_{adh}$ shows a variation with the strain state of the order of a few meV per per cent strain, of the same order of its numerical accuracy. It must also be noticed that, in contrast to the behaviour of the total energy in the free-standing case, compressive and tensile strain can have qualitatively different effects on $E_{adh}$: in general, compression favors out-of-plane distortions that enhance bonding with the substrate, whereas tension tends to keep the overlayer flat (if not broken), thereby reducing its ability to establish strong interfacial bonds.

\subsection*{Registry with the substrate}

Because of the lattice mismatch, the registry of the overlayer with the substrate changes within the supercells. However, to get some insight into the bonding mechanism of hBN to Pt,
we evaluate six distinct (1$\times$1) configurations of a highly stretched hBN on a (1$\times$1) Pt(111) slab to determine their relative stability. The energetically favourable arrangement involves N atoms occupying the top sites of Pt, while B atoms are settled at hollow hcp or fcc sites. This structural preference, corresponding to a 0.02-0.03 eV stronger $|E_{adh}|$ per BN pair, is consistent with earlier DFT findings by  Grad et al.~\cite{Grad_2003} for hBN on commensurate substrates like Ni(111), suggesting the universality of this bonding configuration across TMs substrates.

\subsection*{Orientation with respect to the substrate}

\begin{figure*}[htb!]
    \centering
    \includegraphics[scale=0.5]{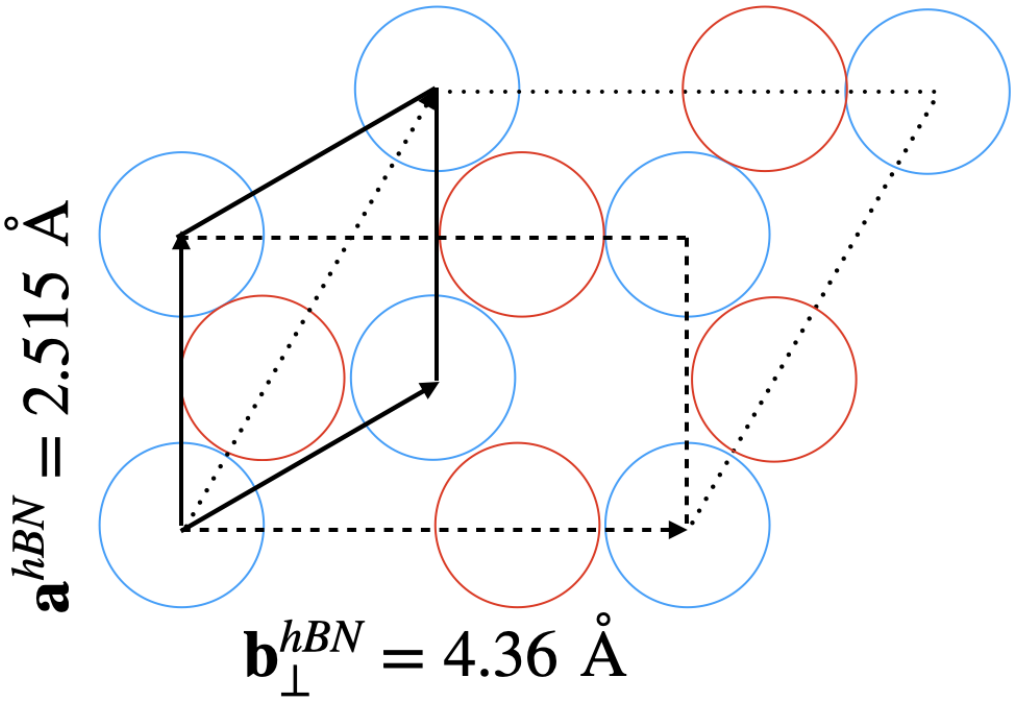}
    \caption{Three different hBN unit cells: primitive unit hexagonal cell (solid lines), rectangular unit cell (dashed lines), hexagonal unit cell with primitive vectors along the armchair directions (dotted lines), containing respectively 1, 2, 3 BN pairs. }
    \label{fig:S5}
\end{figure*}

We complete the investigation of hBN on the Pt(111) surface by studying whether the epitaxially aligned configuration of hBN on Pt(111) is favored with respect to other possible orientations.
From the literature, it emerges that the zigzag direction of a honeycomb overlayer often aligns with a substrate close-packed direction rather than the armchair direction doing so, but this is not universal. The existence of a preferential orientation could be dictated by a coincidence geometry in case of lattice-matched systems, such as graphene on Ni(111)~\cite{Africh2025}, but it is impossible to establish a general rule in case of mismatch.
Focusing primarily on edge-driven orientation selection during island nucleation and growth, Dong et al.~\cite{DongNatureComm2020} show that the propagating edge of a 2D material tends to align along a high-symmetry direction of the substrate and, in particular, for hBN on lattice-matched Cu, the zigzag edges of a nanoribbon are aligned with the substrate close-packed directions $\langle110\rangle$. Novaco and McTague~\cite{Novaco} proposed a theory that predicts an optimum rotation angle depending on mismatch and elastic constant. The application of these theories to our system requires some caution, since the former considers the edges of the overlayer and the latter considers flat adsorbates over rigid substrates.
We therefore perform a direct investigation by simulating the hBN/Pt(111) interface with another orientation.
A rotation of $30^{\circ}$ (labelled with R30) brings the hBN armchair directions aligned with the $\langle110\rangle$ directions of Pt(111).  A structural mismatch remains also in this case, since the periodicity along the armchair directions is 4.36 {\AA}, not commensurate with  the substrate (Fig.~\ref{fig:S5}). Therefore, also for this orientation, we calculate the smallest possible supercells with a strain applied to hBN within 5\% and less than 150 BN pairs, and we report the structural parameters and $E_{adh}$ for some selected configurations in the lowest panel of Tab.~\ref{Tab:2}.
To compare the results for R0 and R30, configurations with similar strain on hBN should be considered; therefore, for instance, the (8$\times$8)hBN$_{hex~armchair}$/((7$\times$7)Pt of R0 could be compared with the (2$\times$2)hBN$_{hex~armchair}$/((3$\times$3)Pt of R30   (Fig. \ref{fig:S3}). In R30, the binding is slightly weaker than in R0, with a difference of only 0.04 eV per BN pair. The difference is nevertheless larger than the numerical accuracy. 

Our DFT results, together with those of Dong et al. \cite{DongNatureComm2020}, which could be applied to the first stages of growth of hBN on the c-(331)Pt support, could explain the experimental evidence of the hBN zigzag alignment along the close-packed steps of vicinal surfaces and the $\langle110\rangle$ directions on the terraces. 

\section*{Models of Pt vicinals}

\begin{figure*}[htb!]
    \centering
    \includegraphics[scale=0.4]{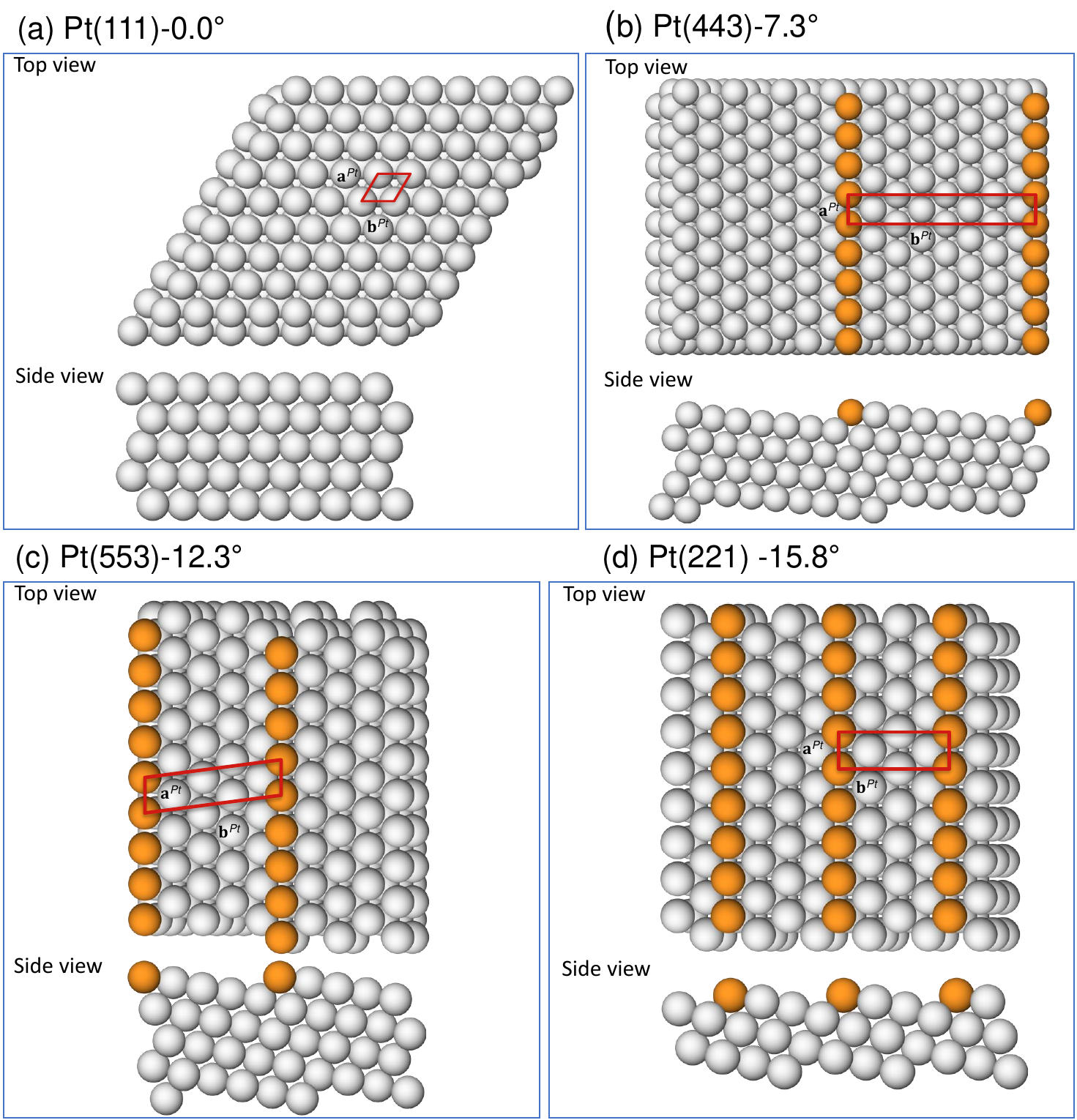}
    \caption{The top view and side view of Pt surfaces at different orientations: (a) (111)-0$^{\circ}$, (b) (443)-7.3$^{\circ}$, (c) (553)-12.3$^{\circ}$, and (d) (221)-15.8$^{\circ}$, with the periodicity vectors of ${\bf a}^{Pt}$ and ${\bf b}^{Pt}$ indicated on each surface.}
    \label{fig:6}
\end{figure*}

\begin{figure*}[htb!]
    \centering
    \includegraphics[scale=0.4]{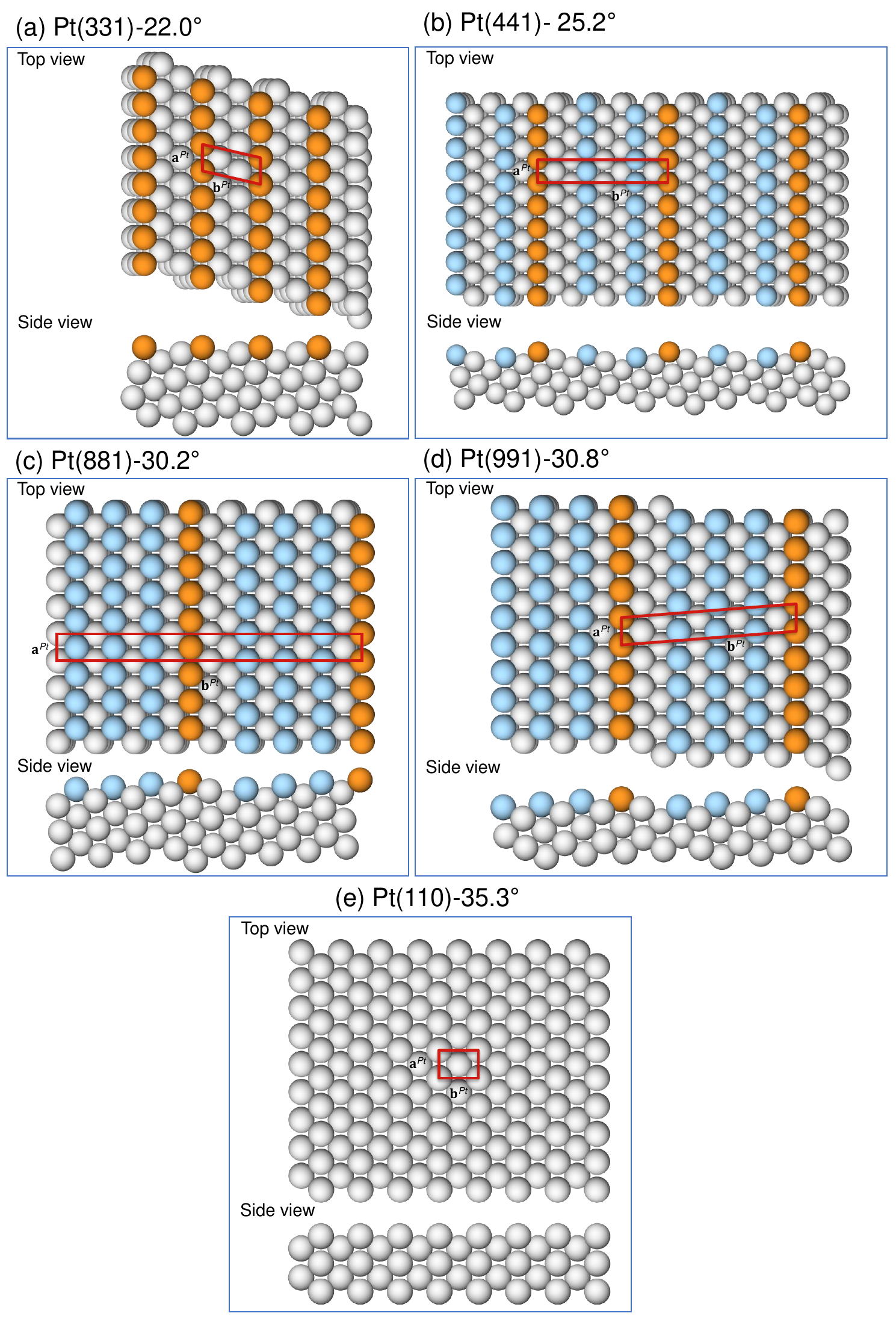}
    \caption{Top and side views of various vicinal Pt surfaces: (a) Pt(331)-$\alpha$=22.00$^{\circ}$, (b) (441)-$\alpha$=25.2$^{\circ}$, (c) (881)-$\alpha$=30.2$^{\circ}$, (d) (991)-$\alpha$=30.8$^{\circ}$ and (e) (110)-$\alpha$=35.3$^{\circ}$. The red surface unit cells with periodicity vectors ${\bf a}^{Pt}$ and ${\bf b}^{Pt}$ are labelled accordingly for each vicinal surface.}
    \label{fig:7}
\end{figure*}

Figs.~\ref{fig:6} and \ref{fig:7} illustrate the atomic structures of different vicinal surfaces of Pt, with increasing miscut angle $\alpha$ with respect to the (111) surface. For the vicinal surface $(hkl)$, the miscut angle is given by $\cos(\alpha) =  \frac{h+k+l}{\sqrt{3}\sqrt{h^2+k^2+l^2}}$. Each panel includes both top and side views of the surface, where surface unit cells are highlighted in red, defined by the periodicity vectors ${\bf a}^{Pt}$ and ${\bf b}^{Pt}$. In the side views, which are taken in the direction perpendicular to ${\bf b}^{Pt}$, only the uppermost Pt layers are represented, whereas more layers are included in the calculations.

In Fig. \ref{fig:6}, panel (a) represents the Pt(111) surface, while all the other panels in Figs. \ref{fig:6} and \ref{fig:7}  illustrate vicinal surfaces to (111) with increasing miscut angle $\alpha$. 
Along the step direction $[1\Bar{1}0]$, the magnitude of the periodicity vector ${\bf a}_{\parallel}^{Pt}$ remains the same, equal to the in-plane lattice parameter ${\bf a}_{(111)}^{Pt}$ across all surfaces, as this direction is preserved due to the crystal being shaped perpendicular to $[1\Bar{1}0]$. However, the other primitive vector ${\bf b}^{Pt}$ forms with ${\bf a}_{\parallel}^{Pt}$ an angle of 90$^\circ$ or slightly different, since, according to the specific Miller indices $(hkl)$, adjacent steps are perfectly aligned or shifted one relative to each other by ${\bf a}_{\parallel}^{Pt}$/2.  
The magnitude of \textbf{b}$^{Pt}$, as well as the periodicity $b_\perp^{Pt}$ perpendicular to the step direction, varies between the vicinal surfaces and reflects the variation of the terrace width with the miscut angle. This variation, not regular,  is evident in the elongated and distorted surface unit cells shown in Figs. \ref{fig:6} and \ref{fig:7}, highlighting the complexity of the morphology of the vicinal surfaces to (111), discussed in detail in the following.

In Fig. \ref{fig:6}, besides the flat (111) surface,  the other panels represent (b) (443)-$\alpha$=7.33$^{\circ}$, (c) (553)-$\alpha$=12.27$^{\circ}$, and (d) (221)-$\alpha$=16.00$^{\circ}$ vicinals.
As $\alpha$ increases from (443) to (221), the surfaces become progressively more stepped, introducing terraces separated by monatomic steps. For the vicinal surfaces like these that can be written as $(m+1,m+1,m-1)$, $m$ is the number of entire Pt rows of a terrace: $m$=
7, 4, 3 respectively for (443), equivalent to (886); (553); (221), equivalent to (442), corresponding to lateral spans of about 7--15 \AA.
If $m$ is odd, like for (443) and (221), ${\bf a}_{\parallel}^{Pt}$ and ${\bf b}^{Pt}$ are perpendicular, and the unit cell is rectangular~\cite{note1}. More generally, for vicinal surface $(hkl)$ of an $fcc$ crystals with B-type steps, where $h$=$k$,  ${\bf a}_{\parallel}^{Pt}$ and ${\bf b}^{Pt}$ are perpendicular if $h$ is even.

Fig.~\ref{fig:7} presents other Pt vicinal surfaces, including (331), (441), (881), (991) and (110), corresponding to $\alpha$= 22.0$^{\circ}$, 25.2$^{\circ}$, 30.2$^{\circ}$, 30.8$^{\circ}$ and 35.3$^{\circ}$, respectively. Panel (a) shows the Pt(331) surface, which is located in the centre of the curved Pt(331) crystal, features two atom-wide terraces, and is the most densely stepped surface. Note that as the vicinal angle increases from zero until it reaches the (331) surface, the step density correspondingly becomes higher, leading to more compact and narrow terraces. Beyond this vicinal surface, the terraces become wider, as seen for (441)-25.80$^{\circ}$. In panel (b), the Pt(441) surface displays a periodic array of alternating terraces one- and two-atom wide. This is indicative of the transitional regime, where the (111) terraces are disappearing, and the (110) terraces are emerging. Panels (c) and (d) illustrate surfaces where (110) terraces dominate, with widths ranging between 4 and 12~\AA. These configurations highlight the transition from the (111) terraces to (110) as the vicinal angle increases beyond 25.2$^{\circ}$. 
Panel (e) presents the Pt(110)-35.10$^{\circ}$ surface, characterised by its most compact configuration, with terraces that are only one atom wide. 

Tab.~\ref{Tab:2} summarizes the crystallographic parameters of the vicinals considered in this work and the corresponding DFT-calculated free energy $\gamma$ (defined in the Main Text).

\begin{table*}[htb!]
\begin{center}
\scalebox{0.9}{\begin{tabular}{|c|l|l|l|r|l|l|}
\hline
vicinal & $\alpha$ & ${\bf b}^{Pt}$ ($a_{bulk}$ units) & angle with ${\bf a}^{Pt}$ & $b_\perp^{Pt}$({\AA}) & structure of the largest terraces&$\gamma$ (eV/{\AA)}$^2$\\
\hline
(111) &  0$^\circ$& &  &&&0.136\\
\hline
(443) &  7.33$^\circ$&[3 3 $\bar{8}$] / 2  &  90$^\circ$      &10.77&(111), 7 atomic rows&0.143\\
(553) & 12.27$^\circ$&[2 1 $\bar{5}$] / 2  &  82.58$^\circ$ &10.66&(111), 4 atomic rows&0.149\\
(221) & 15.79$^\circ$&[1 1 $\bar{4}$] / 2  &  90$^\circ$      &8.32&(111), 3 atomic rows&0.151\\
(331) & 22.00$^\circ$&[1 0 $\bar{3}$] / 2  &  77.08$^\circ$ &6.04 &(111), 2 atomic rows&0.160\\
(441) & 25.24$^\circ$&[1 1 $\bar{8}$] / 2  &  90$^\circ$      &15.94&(111), 2 atomic rows on avg &0.158\\
&&&&&OR (110), 1.5 atomic rows on avg &\\
(881) & 30.21$^\circ$&[1 1 $\bar{16}$] / 2 &  90$^\circ$     &31.51&(110), 3.5 atomic rows on avg &0.160\\
(991) & 30.77$^\circ$&[1 0 $\bar{9}$] / 2   &  85.52$^\circ$ &17.72&(110), 4 atomic rows&0.160\\
(110) & 36.26$^\circ$&[0 0 $\bar{1}$]        & 90$^\circ$       &3.92& &0.165\\
\hline
\end{tabular}}
\caption{Structural parameters and surface free energy of  Pt(111) and its vicinals ordered according to the miscut angle: for each vicinal, it is reported the miscut angle $\alpha$, the shorter surface primitive vector  ${\bf b}^{Pt}$ in units of the bulk lattice parameter, the angle with the other surface primitive vector ${\bf a}^{Pt}=\frac{1}{2}a_{bulk}[1\bar{1}0]$, which is common to all these surfaces and parallel to the steps, and, for the specific case of Pt, the step-to-step distance ${b}_\perp^{Pt}$.}
\label{Tab:2}
\end{center}
\end{table*}

\clearpage

\section*{Models of hBN on Pt vicinals}

\medskip

Building commensurate cells between hBN and Pt vicinals with the additional constraint of zero or almost zero misalignment or \emph{tilting} angle $\theta$ with the step edges is rather challenging and, to our knowledge, has not been afforded up to now. Different strategies could be followed. Focusing on   $\theta$=0, a possibility
is to consider as the primitive unit cell of hBN a rectangular cell with ${\bf a}_{\parallel}^{hBN}={\bf a}^{hBN}$=2.515 {\AA} (one zig-zag unit) and ${\bf b}_{\perp}^{hBN}$=4.36 {\AA} (one arm-chair unit) (see Fig. \ref{fig:S5}). Therefore, matching hBN with the substrate for $\theta$=0 would require a cell with 10 hBN zig-zag units and 9 Pt atoms along the step, with a strain on hBN in this direction equal to  $-$0.9\%. A better alternative could be obtained with 11 hBN zig-zag units and 10 Pt atoms along the step, with a strain on hBN of only +0.1\% (see also Tab.~\ref{Tab:1}). We keep the former condition to reduce the computational complexity.
Matching the overlayer with the substrate in the other direction could be obtained with a multiple of ${\bf b}^{Pt}$  commensurated with a multiple of ${\bf b}_{\perp}^{hBN}$, which is a strong condition, even within a certain tolerance of a few per cent.

In Tab \ref{Tab:3} we report for each vicinal the description of one or more possible supercells obeying these conditions and with a strain on hBN smaller than 5\% in absolute value. The different facets are listed in order of increasing miscut angle $\alpha$. For each vicinal, we select the "optimal" supercell considering a compromise between the computational effort and the strain state, and perform DFT calculations.  In general, for a similar absolute value of the strain, a compressive (negative) rather than tensile (positive) strain should be preferred in our opinion, since a tensile strain tends to keep the overlayer flat, reducing the possibility of interfacial
bonds. For vicinals described by oblique surface primitive unit cells, with a {\bf b}$^{Pt}$ vector not orthogonal to the step edge, it would be preferable to consider an even number of unit cells in order to use rectangular supercells, avoiding a {\em {shear}} strain on hBN.

Interface models could also be built following the general strategy commonly used for moir\'e structures, starting from the primitive unit cells of the two superimposed lattices. In our case, this means to consider multiples of the hBN hexagonal primitive unit cell, with ${\bf a}=N_a {\bf a}^{hBN}+N_b{\bf b}^{hBN}$ and ${\bf b}=M_a{\bf a}^{hBN}+M_b{\bf b}^{hBN}$, and match them with a Pt vicinal supercell with ${\bf a}=P_a {\bf a}^{Pt}$ and ${\bf b}=P_a {\bf b}^{Pt}$. This approach in principle does not guarantee either the alignment of the hBN zigzag directions with the Pt $\langle 110\rangle$ directions or the absence of a {\em {shear}} strain on hBN. However, suitable models can also be obtained with this approach, with zero tilting angle if $N_b=0$ and small {\em shear} strain, as we report for (331) and (881).

\begin{figure*}[htb!]
    \centering
    \includegraphics[scale=0.5]{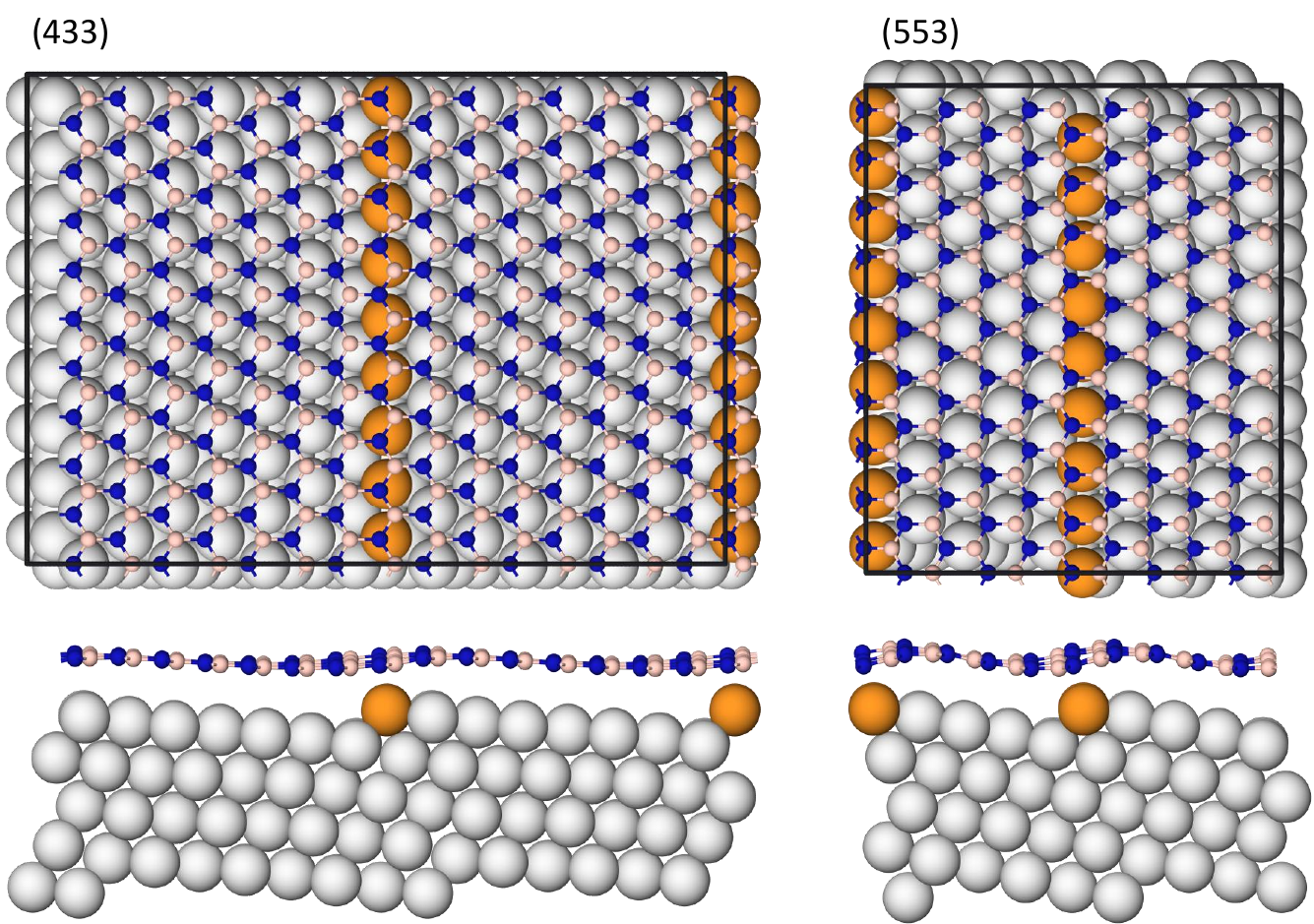}
    \caption{Top and side view of the DFT-optimized model for hBN on (443) and (553) Pt vicinals. }
    \label{fig:S8}
\end{figure*}

\begin{figure*}[htb!]
    \centering
    \includegraphics[scale=0.5]{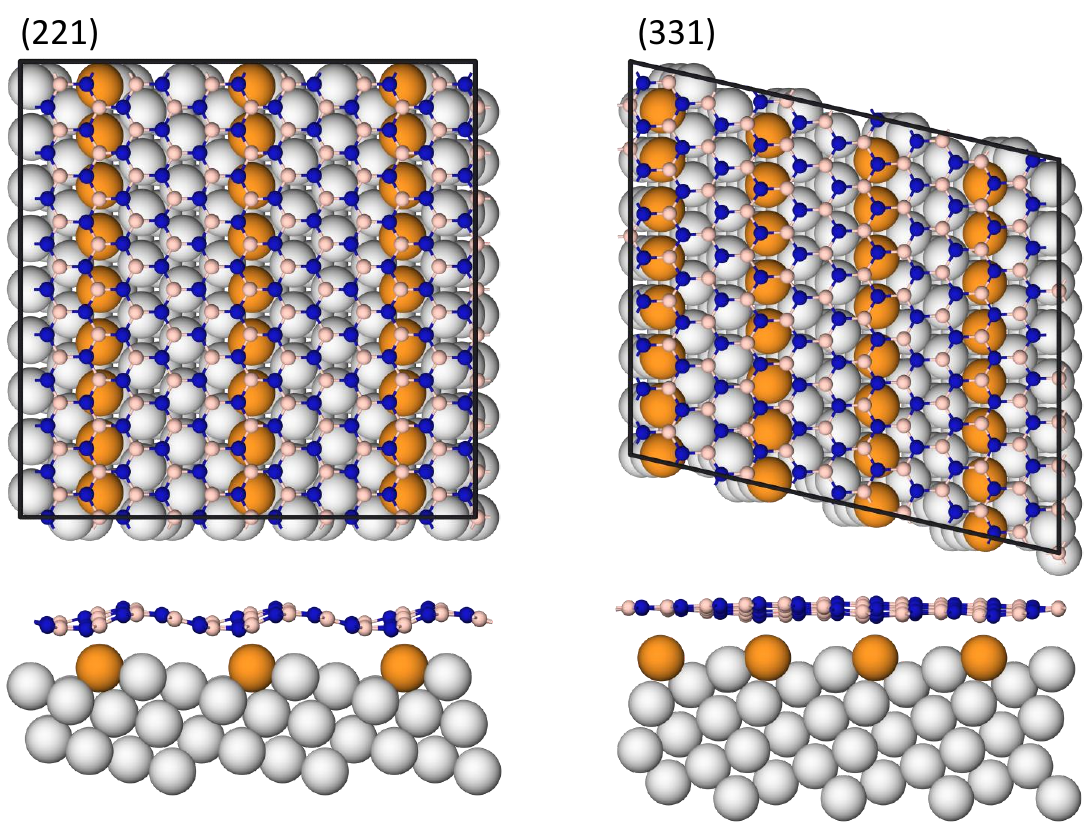}
    \caption{Top and side view of the DFT-optimized model for hBN on (221) and (331) Pt vicinals. }
    \label{fig:S9}
\end{figure*}

\begin{figure*}[htb!]
    \centering
    \includegraphics[scale=0.5]{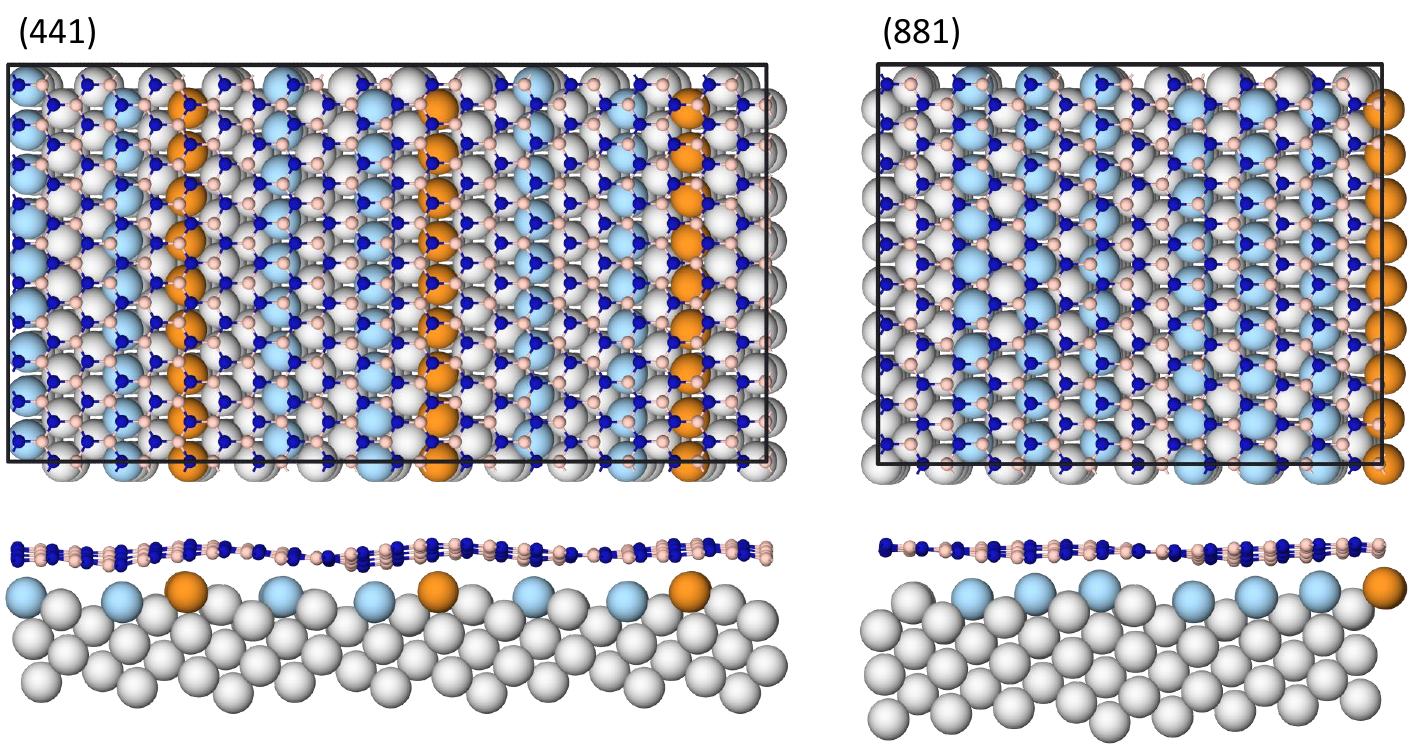}
    \caption{Top and side view of the DFT-optimized model for hBN on (441) and (881) Pt vicinals. }
    \label{fig:S10}
\end{figure*}

\begin{figure*}[htb!]
    \centering
    \includegraphics[scale=0.5]{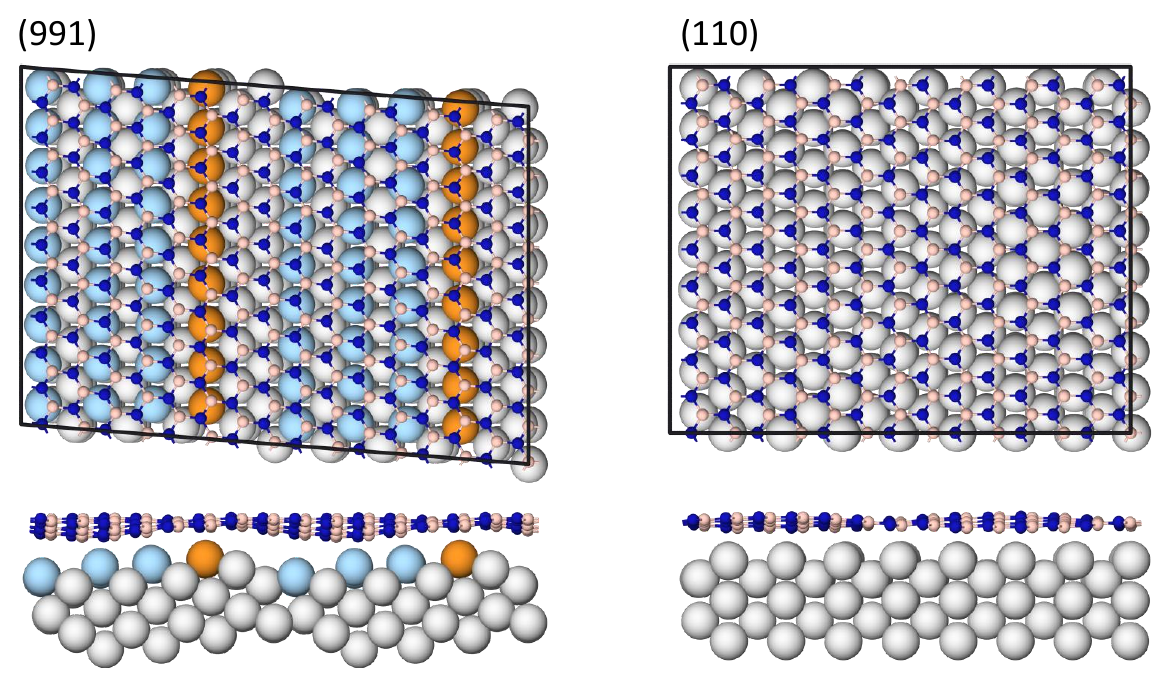}
    \caption{Top and side view of the DFT-optimized model for hBN on (991) and (110) Pt vicinals. }
    \label{fig:S11}
\end{figure*}

Figs.~\ref{fig:S8}-\ref{fig:S11} show our best DFT-optimized models for hBN on Pt vicinals.

\clearpage
\begin{table*}[htb!]
\begin{center}
\scalebox{0.7}{\begin{tabular}{|c|c|c|c|c|c|c|l|}
\hline
no. Pt cells& no. hBN rectangular cells& $b_\perp^{Pt}$({\AA}) & $b^{hBN}$({\AA}) & strain on $b^{hBN}$ (\%)&  no.  BN pairs&E$_{adh}$(eV)& notes\\
\hline\hline
\multicolumn{8}{|l|}{(443), rectangular; b(Pt unit cell) = 17.77 {\AA}} \\
\hline (9$\times$1) &  (10$\times$4) &  17.77 &  17.44 &  1.89 &80&\hey{-0.18}& \\
\hline\hline
\multicolumn{8}{|l|}{(553), oblique; b$_\perp$(Pt unit cell) = 10.66 {\AA} (b=10.75 {\AA}, $\theta$=82.58)}\\ 
\hline (9$\times$2) &  (10$\times$5)  &  21.32& 21.80  &  -2.20 &100&\hey{-0.20}& \\
 (9$\times$3)&  (10$\times$7)  &  31.98& 30.52  &   4.78 &140&&with {\em shear} strain\\
\hline\hline
\multicolumn{8}{|l|}{(221), rectangular; b(Pt unit cell) =  8.32 {\AA}}\\ 
 \hline (9$\times$1)&  (10$\times$2)  &   8.32&  8.72  &  -4.59 &40&\textcolor{green}{\bf -0.23}& high compressive strain: corrugation 1.17 {\AA} \\
 (9$\times$6) & (10$\times$11)  &  49.92& 47.96  &   4.09 &220&&\\
\hline\hline
\multicolumn{8}{|l|}{(331), oblique; b$_\perp$(Pt unit cell) =  6.04 {\AA} (b=6.20 {\AA}, $\theta$=77.08)} \\ 
 \hline (9$\times$3) &  (10$\times$4)  &  18.12& 17.44  &   3.90 &80&&with {\em shear} strain\\
 (9$\times$5)&  (10$\times$7)  &  30.20& 30.52  &  -1.05 &140&&with {\em shear} strain\\
 (9$\times$7)& (10$\times$10)  &  42.28& 43.60  &  -3.03 &200&&with {\em shear} strain\\
 (9$\times$8)& (10$\times$11)  &  48.32& 47.96  &   0.75 &220&& optimal supercell:  rectangular, smallest strain\\
 (8$\times$4)& (9$\times$11) primitive cells &24.16&&&99&\hey{-0.19} & {\em shear} angle 1.78$^\circ$(*)\\
\hline\hline
\multicolumn{8}{|l|}{(441), rectangular; b(Pt unit cell) = 15.94 {\AA}}\\
 \hline (9$\times$2)&  (10$\times$7)  &  31.88& 30.52  &   4.46&140&\textcolor{green}{\bf -0.20}&high tensile strain\\
 (9$\times$3)& (10$\times$11)  &  47.82& 47.96  &  -0.29 &220&\textcolor{green}{\bf -0.19}&\\
\hline\hline
\multicolumn{8}{|l|}{(881), rectangular; b(Pt unit cell) = 31.51 {\AA}}\\  
\hline (9$\times$1)&  (10$\times$7)  &  31.51& 30.52  &   3.24 &140&\textcolor{green}{\bf -0.20}&\\
\hline (8$\times$1) &  (9$\times$14) primitive cells &  31.51&   &    &126&\textcolor{green}{\bf -0.21}& {\em shear} angle 0.32$^\circ$ (**)\\
\hline\hline
\multicolumn{8}{|l|}{(991), oblique; b$_\perp$(Pt unit cell) = 17.72 {\AA} (b=17.77 {\AA}, $\theta$=85.52)} \\ 
 \hline (9$\times$1)&  (10$\times$4)  &  17.72& 17.44  &   1.61 &80&\hey{-0.20}& {\em shear} angle  4.48$^\circ$\\
 (9$\times$2)&  (10$\times$8)  &  35.44& 34.88  &   1.61 &160&&same ratio as above, but rectangular supercell\\ 
\hline\hline
\multicolumn{8}{|l|}{(110), rectangular; b(Pt unit cell) =  3.92 {\AA}}\\ 
\hline (9$\times$7)&   (10$\times$6)   &  27.44& 26.16  &   4.89&120&&\\
 (9$\times$8)&   (10$\times$7)   &  31.36& 30.52  &   2.75&140&\textcolor{green}{\bf -0.22}&\\
 (9$\times$9)&   (10$\times$8)   &  35.28& 34.88  &   1.15&160&&\\
 (9$\times$10)&  (10$\times$9)   &  39.20& 39.24  &  -0.10 &180&&\\
 (9$\times$11)&  (10$\times$10)  &  43.12& 43.60  &  -1.10&200&&\\
 \hline
\end{tabular}}
\caption{Structural parameters of supercells describing hBN overlayer on vicinals of Pt(111) ordered according to the miscut angle: for each vicinal, the structure of its primitive surface unit cell is described (rectangular or oblique, the value of $b$ and its projection in the direction perpendicular to the step, corresponding to step-to-step distance); for each supercell, we report: the number of repeated Pt vicinal and hBN rectangular unit cells, their respective dimensions in the direction perpendicular to the step edgs, the strain applied to hBN to match the Pt supercell in that direction, the number of BN pairs, and, for the selected structures, the adhesion energy for hBN formula unit. Since, if not stated otherwise, along the step direction 10 hBN unit cells are accommodated on 9 Pt unit cells, the strain in the zigzag direction of hBN is always $-$0.9\%. The last column reports further details; (*) and (**) label supercells built with a different strategy, starting from primitive and non-rectangular hBN cells. In particular, the lattice vectors of the quasi-commensurate hBN supercells are in the two cases: (*) ${\bf a}=9 {\bf a}^{hBN}; {\bf b}=8{\bf a}^{hBN}+11{\bf b}^{hBN}$; (**) ${\bf a}=9 {\bf a}^{hBN}; {\bf b}=8{\bf a}^{hBN}+14{\bf b}^{hBN}$. The adhesion energy values are in green/red, indicating the experimentally stable/unstable facets after hBN growth. 
}
\label{Tab:3}
\end{center}
\end{table*}

\clearpage
\bibliography{rsc-SM} 
\bibliographystyle{elsarticle-num}